\documentclass[twocolumn]{aastex63}
\usepackage{hyperref}
\usepackage[colorinlistoftodos]{todonotes}
\usepackage{amsmath}
\shorttitle{Masses from low-S/N TTVs}

\begin{document}
\title{Mass Upper Bounds for Over 50 Kepler Planets Using Low-S/N Transit Timing Variations}

\correspondingauthor{Jared Siegel}
\email{siegeljc@uchicago.edu}

\author[0000-0002-9337-0902]{Jared C. Siegel}
\affiliation{Department of Astronomy and Astrophysics, University of Chicago, Chicago, IL 60637, USA}

\author[0000-0003-0638-3455]{Leslie A. Rogers}
\affiliation{Department of Astronomy and Astrophysics, University of Chicago, Chicago, IL 60637, USA}

\begin{abstract}
Prospects for expanding the available mass measurements of the Kepler sample are limited. Planet masses have typically been inferred via radial velocity (RV) measurements of the host star or time-series modeling of transit timing variations (TTVs) in multiplanet systems; however, the majority of Kepler hosts are too dim for RV follow-up, and only a select number of systems have strong enough TTVs for time-series modeling. Here, we develop a method of constraining planet mass in multiplanet systems using low signal-to-noise ratio (S/N) TTVs. For a sample of 175 planets in 79 multiplanet systems from the California–Kepler Survey, we infer posteriors on planet mass using publicly available TTV time-series from Kepler. For 53 planets ($>30\%$ of our sample), low-S/N TTVs yield informative upper bounds on planet mass, i.e., the mass constraint strongly deviates from the prior on mass and yields a physically reasonable bulk composition. For 25 small planets, low-S/N TTVs favor volatile-rich compositions. Where available, low-S/N TTV-based mass constraints are consistent with RV-derived masses. TTV time-series are publicly available for each Kepler planet, and the compactness of Kepler systems makes TTV-based constraints informative for a substantial fraction of multiplanet systems. Leveraging low-S/N TTVs offers a valuable path toward increasing the available mass constraints of the Kepler sample. 
\end{abstract}


\section{Introduction}
\label{sec:intro}

Through the detection of thousands of transiting planets---coupled with a well-studied completeness function---the Kepler mission has facilitated detailed investigations of planetary demographics. Using planet radius and period constraints from the Kepler data, prior studies demonstrated the prevalence of small planets \citep[e.g.,][]{Howard2012,Petigura2013,Dressing2015}, constrained the multiplicity distribution of planetary systems \citep[e.g.,][]{Fang2012,Zhu2018}, identified the Kepler dichotomy \citep[e.g.,][]{Lissauer2011,Johansen2012,Millholland2021}, and revealed a gap in the radius distribution of planets between $1.5$ and $2.0$~$R_{\oplus}$ \citep{Fulton2017}.

While demographics studies based upon transit surveys often consider planets in terms of radius and period, planet mass is critical to understanding planetary composition, formation, and evolution. Recently, \cite{Neil2020} explored the joint mass-radius-period distribution of planets via hierarchical Bayesian mixture models and concluded models with gaseous, evaporated core, and intrinsically rocky populations were preferred by the data. However, such studies are limited by the available mass measurements of the Kepler sample. Without mass estimates, degeneracies between planet radius and composition persist for super-Earths and sub-Neptunes \citep[e.g.,][]{Valencia2007,Adams2008,Rogers2010}. To break these degeneracies and advance studies of planetary demographics, there is considerable demand for increased mass measurements of the Kepler sample.

For planets in the Kepler sample, masses are typically inferred via radial velocity (RV) measurements of the host star or transit timing variations (TTVs). TTVs are shifts in a planet's mid-transit time relative to strict Keplerian periodicity \citep{Agol2005,Holman2005}. However, most Kepler host stars are too faint to be accessible to current or next-generation RV facilities. Prospects for expanding the sample of Kepler planets with RV-derived mass constraints therefore remain limited. 

Contrary to RV measurements, TTV time-series are extracted directly from Kepler light-curves and are available for every Kepler Object of Interest \citep[KOI; see the TTV catalogs of][]{Rowe2015,Holczer2016}. In addition to stellar activity and instrumental noise, structured variations in a planet's measured mid-transit times can arise from dynamical interactions between a planet and other objects in the system. Via numerical integration \citep{Deck2014}, first-order in eccentricity and mass analytic solutions \citep{Agol2016}, or a linear combination of basis functions \citep{Hadden2016,Linial2018,Hadden2019} prior studies have constrained the masses and eccentricities of planets in large samples of multiplanet systems by TTV time-series modeling \citep{Xie2013,Hadden2014,Xie2014,JontofHutter2016,Hadden2017}; analytic light-curve modeling has also been applied to characterize planet mass and orbital elements \citep{Judkovsky2022}. Given the accessibly of TTVs, TTV-based mass constraints have been a vital source of mass measurements for the Kepler sample.  

Thus far, masses have been constrained in $>50$~multiplanet systems using TTVs. These studies were restricted to systems in which the TTVs are particularly strong. Of the 145 planets in 55 systems characterized by \cite{Hadden2017}, the average excess scatter ratio is $>12$, where the excess scatter ratio is defined as the standard deviation of the observed TTVs divided by the mean error of the mid-transit time measurements; for the remaining systems, the mean excess scatter ratio is $\sim2$. For this study, TTV time-series with an excess scatter ratio less than $5$ are considered low-S/N TTVs. While systems with low-S/N TTVs have been bypassed by prior studies, such systems represent the vast majority of multiplanet systems. More than $700$ systems in the \cite{Holczer2016} catalog have reported TTV measurements but lack TTV-based mass constraints. Mass constraints derived from low-S/N TTVs would significantly increase the available mass information for the Kepler sample.

In this paper, we develop a method of inferring conservative constraints on planet mass using low-S/N TTV data. While low-S/N TTVs are generally inaccessible to time-series modeling, a KOI's TTVs can be aggregated into a higher signal summary statistic (e.g., variance). Aggregating TTVs into a summary statistic neglects time-correlated structure and is therefore only reasonable for low-S/N data. Through a series of simplifying approximations and by decomposing the observed TTVs into three independent sources (noise from measurement uncertainties, stellar noise, and planet-planet perturbations), we derive a likelihood function for the observed TTV variances in a system given a vector of system properties (e.g., masses, orbital periods, eccentricities, and phases). Posteriors on planet mass are then inferable via Monte Carlo sampling.  

This paper is organized as follows. In Section~\ref{sec:sample}, we describe our sample of Kepler systems. Then, in Section~\ref{sec:methods}, we outline our mass inference methods. The results from applying our methods to Kepler systems are reported in Section~\ref{sec:results} and discussed in Section~\ref{sec:disc}. We conclude in Section~\ref{sec:conc}. 

\section{Sample of Kepler Systems}
\label{sec:sample}

For this study, we consider the TTV catalog of \cite{Holczer2016}. This catalog includes $>2500$ KOI and considers all 17 quarters of data. In addition to removing false-alarm candidates, \cite{Holczer2016} selected KOIs with a transit S/N $>7.1$, a transit depth $<10\%$ (to avoid eclipsing binaries), and an orbital period $<300$~days. Outlier TTV measurements are flagged and removed \citep[for discussion of outliers, see Section 2.4 of][]{Holczer2016}.

\begin{figure}[t]
\gridline{\fig{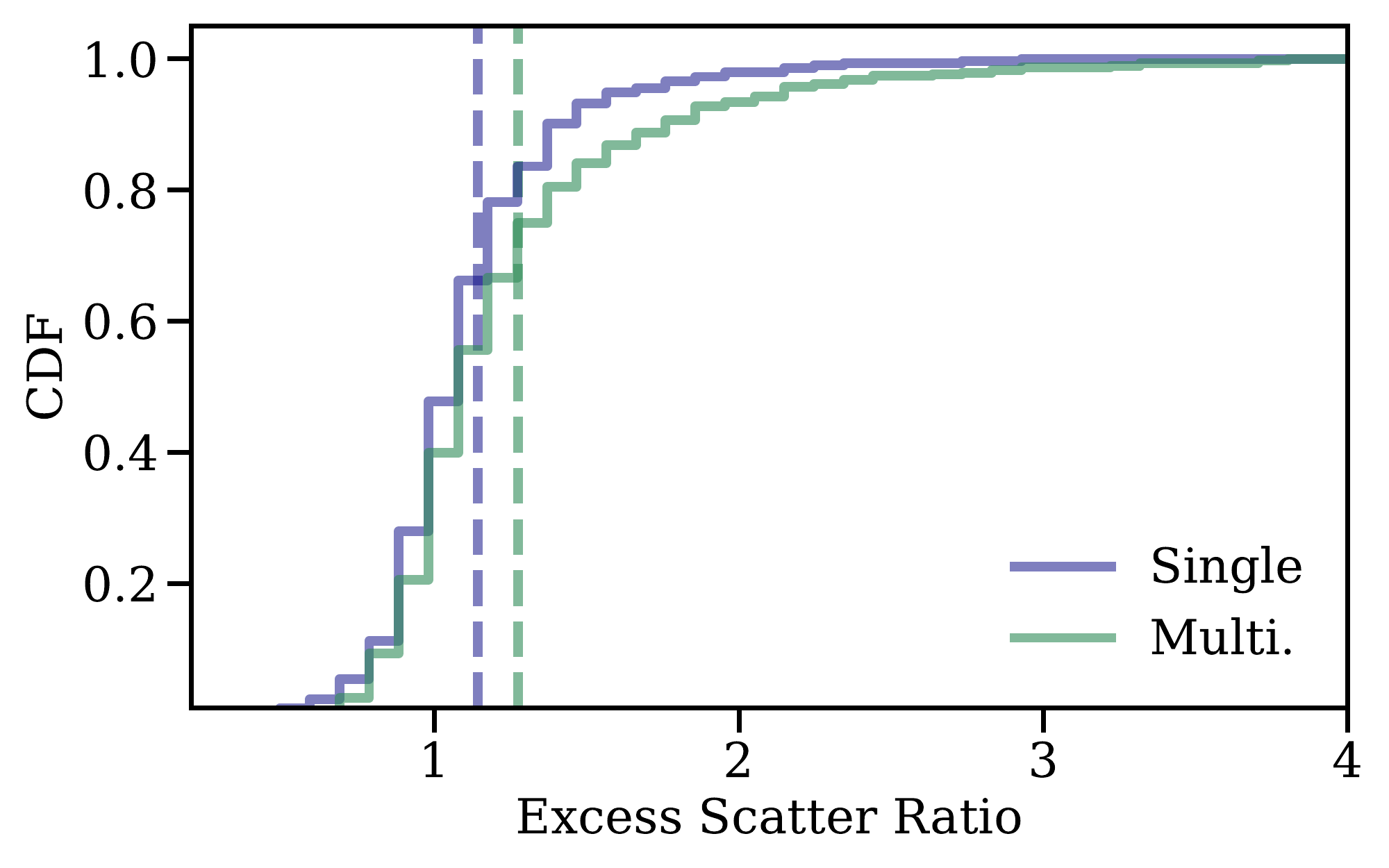}{\columnwidth}{}}
\caption{ The excess scatter ratio distributions for systems with a single transiting planet known to date (blue) and systems with multiple known transiting planets (green) both peak near $1$ with a skew toward higher values. The mean excess scatter ratio of the planets in single planet systems is shown as a vertical blue dashed line, and the mean excess scatter ratio of the planets in multiplanet systems is shown as a vertical green dashed line.}
\label{fig:S/N}
\end{figure}

As defined in Section~\ref{sec:intro}, the excess scatter ratio for a given KOI is the standard deviation of its TTV time-series divided by the mean measurement uncertainty of its transit times from \cite{Holczer2016}. For both single and multiplanet systems, the distribution of excess scatter ratio peaks near $1$ and is skewed toward higher values (Figure~\ref{fig:S/N}). Throughout this study, we consider KOIs with low-S/N TTVs, defined as an excess scatter ratio less than $5$. 

Although qualitatively useful, the excess scatter ratio is a nonrigorous summary of a TTV time-series, because it does not depend on the number of observed transits. In Section~\ref{sec:methods}---where we derive the likelihood function for TTV variances in a system for a given set of planet properties (e.g., masses, orbital periods, eccentricities, and  phases)---the full dependence on the number of observed transits is considered.

In the interest of leveraging our results to inform demographics studies of the Kepler sample, we further select KOIs included in the California–Kepler Survey (CKS). Via high-resolution spectroscopic follow-up of host stars and cross-matching with Gaia, the CKS sample offers high-precision radius measurements for transiting planets \citep{Johnson2017,Petigura2017,Fulton2018}. The CKS sample has also been targeted for RV mass measurement follow-up, allowing for comparison between our methods and independent mass constraints. KOIs in the CKS sample are further filtered following \cite{Fulton2018} and \cite{Neil2020}, e.g., cuts on magnitude, stellar temperature, and stellar type. Planet radii are adopted from \cite{Fulton2018}. After filtering, we have 1130 planets in 715 systems. 

Cross-matching the CKS sample with the \cite{Holczer2016} catalog, we have 269 planets in 113 multiplanet systems. To ensure optimal completeness of the TTVs, we next require the TTV catalog includes each confirmed KOI in a system (e.g., a system is rejected if a transiting planet has been discovered or shown to be a false positive since the \cite{Holczer2016} catalog was released). Each KOI must also have $>7$ transits after removing outlier transits. For computational efficiency, the sample is limited to systems with three or fewer transiting planets. 

Finally, systems where a pair of adjacent planets is within 1\% of a first-order mean-motion resonance (e.g., 2/1, 3/2, 4/3, ...) are rejected. These systems likely require tuned orbital elements for long-term stability, which is incompatible with the broad and independent priors on orbital elements adopted herein. For instance, \cite{Gillon2017} initially found that most samples from their fits to TRAPPIST-1 become unstable within 0.5~Myr; however, \cite{Tamayo2017} demonstrated that when the planets were initialized through disk migration, most samples remained stable for 50~Myr. Without proper initialization, samples could become unstable in these near resonant systems and lead to overly stringent mass constraints; see Section~\ref{sec:correlations} for discussion. 

KOI-658 (Kepler-203) is rejected due to the system's dependence on the adopted treatment of stellar noise; see Section~\ref{sec:vstar} for discussion.

Our final sample includes 175 planets in 79 multiplanet systems, and 293 planets in systems with a single detected transiting planet.

\section{Methods}
\label{sec:methods}

In this section, we describe our method of mass inference for low-S/N TTVs. As outlined in Section~\ref{sec:intro}, a given TTV time-series can be aggregated into a higher signal summary statistic: variance. Via a series of simplifying approximations, each KOI's TTV variance can be treated as a sum of three independent components (noise from measurement uncertainties, stellar noise, and planet-planet perturbations). After deriving the likelihood function for the observed TTV variances in a system given a set of system properties (e.g., masses, orbital periods, eccentricities, and  phases), posteriors on planet mass can be inferred via Monte Carlo sampling. 

This method is motivated by the apparent excess of scatter in KOIs' TTV variance relative to their mean measurement uncertainties; see Figure~\ref{fig:S/N}, the distributions of excess scatter ratio for planets in single planet and multiplanet systems. However, as noted in Section~\ref{sec:sample}, the excess scatter ratio is a qualitatively useful, but nonrigorous, summary of TTV time-series, because it does not consider the number of observed transits. A statistical treatment of TTV variances, which considers the number of observed transits, is given below. 

\cite{JontofHutter2016} previously reported a similar excess of TTV scatter in the residuals between the observed transit times and the best-fit TTV models in a sample of 18 planets with strong TTVs. \cite{JontofHutter2016} accounted for this scatter by assuming the mid-transit time measurement uncertainties followed a Student’s $t$-distribution with two degrees of freedom; the model parameter posteriors from the TTV fits of \cite{JontofHutter2016} showed no statistically significant differences between assuming Gaussian uncertainties and assuming the uncertainties followed a Student’s $t$-distribution.

The excess of TTV scatter is more pronounced for planets in systems with multiple transiting planets (see Figure~\ref{fig:S/N}). In the high-S/N regime, \cite{XieWuLithwick2014} identified an analogous relation between the fraction of systems presenting significant TTVs and system multiplicity, using Lomb–Scargle periodograms to identify KOIs with long-term TTV oscillations. Given this multiplicity dependence, the excess scatter is likely due to a combination of stellar noise (e.g., spot crossings, photometric variability from granulation, residuals in long-term light-curve detrending, etc.) or misestimated measurement uncertainties---either of which would affect planets in both single and multiplanet systems---and planet-planet perturbations. The effect of nondetected planets is discussed in Section~\ref{sec:undetected}.

Other than dynamical planet-planet perturbations, multiplanet systems could exhibit a greater excess of TTV scatter due to systematics of the light-curve fitting. In multiplanet systems, the transits of a given planet must be masked out before fitting the transits of the next planet. This reduces the available light-curve baseline and may translate to greater TTV scatter in multiplanet systems. However, \cite{Holczer2016} marked overlapping transits---where the difference between the expected transit times of two planets was smaller than twice the sum of the two planets' transit durations---as outliers, limiting the effect of the light-curve masking on TTV scatter. Planet-planet perturbations are likely the driving factor in the relation between TTV scatter and system multiplicity.

In Section~\ref{sec:decomp}, we first outline the decomposition of TTVs into three independent components: noise from measurement uncertainties, stellar noise, and planet-planet perturbations. In Section~\ref{sec:vstar}, we constrain the distribution of stellar noise for the Kepler sample via hierarchical Bayesian inference of the single-planet systems. Then, in Section \ref{sec:methods_mass}, we discuss our method of generating posteriors on planet mass in multiplanet systems using low-S/N TTVs.

\subsection{Decomposition of low-S/N TTVs}
\label{sec:decomp}

TTVs are shifts of a planet's mid-transit timings ($O-C$) relative to strict Keplerian periodicity. For the $n$th transit of the $i$th planet around the $j$th star, the TTV is defined as
\begin{equation}
    \label{equ:TTV}
    \text{TTV}_{i,j}(n) = T_{{\rm mid},i,j}(n) - P_{i,j}\cdot n -  T_{1,i,j},
\end{equation}
where $P_{i,j}$ is the planet's Keplerian orbital period, $ T_{1,i,j}$ is the initial mid-transit time, and $T_{{\rm mid},i,j}(n)$ is the $n$th mid-transit time; $n=1,\dots,N_{i,j}$, where $N_{i,j}$ is the number of observed transits; $i=1,\dots,N_{ {\rm p},j}$, where $N_{ {\rm p},j}$ is the number of planets in the $j$th system; and $j=1,\dots,N_{\star}$, where $N_{\star}$ is the number of stars in the sample. $P_{i,j}$  is typically determined via a linear regression between $\{n\}$\footnote{ The notation $\{ n  \}$ denotes the series of values $\{ 1,\dots,N_{i,j} \}$.  } and $\{  T_{{\rm mid},i,j}(n) \}$. For notational clarity, the transit numbering assumes there are no gaps in the transit time-series. Eqn.~\ref{equ:TTV} can readily be generalized to include missing transits.

In the limit of infinite-precision measurements, TTVs can be considered as a sum of two independent components,
\begin{equation}
    \text{TTV}_{i,j}(n) = \Phi_{ \text{planets}, i,j }(n) + \Phi_{\star,i,j}(n),
\end{equation}
where $\Phi_{ \text{planets}, i,j }(n)$ is the mid-transit time variation induced by planet-planet perturbations, and $\Phi_{\star,i,j}(n)$ is the mid-transit time variation due to activity on the host star's surface (e.g., spot crossings, photometric variability from granulation, etc.). Prior studies of TTV mass estimation often hinge on modeling $\Phi_{ \text{planets}, i,j }$ as a function of time and either neglect $\Phi_{\star,i,j}$ or statistically account for $\Phi_{\star,i,j}$ as white noise. However, such modeling of $\Phi_{ \text{planets}, i,j }$ relies on high-S/N data. 

To extract mass constraints from low-S/N data, a series of simplifying approximations must be made. First, over an observing baseline sufficiently long relative to the effective period of the TTVs, $\Phi_{ \text{planets}, i,j }$ and $\Phi_{\star,i,j}$ are each assumed to be well characterized by a time-independent population variance, $V_{ \text{planets},i,j}$ and $V_{\star,i,j}$ respectively. Since $\Phi_{ \text{planets}, i,j }$ and $\Phi_{\star,i,j}$ are independent of each other, the population variance of a planet's TTVs in the limit of infinite-precision measurements is
\begin{align}
    \label{equ:var_no_noise}
    V_{i,j} = V_{ \text{planets},i,j} + V_{\star,i,j}.
\end{align}
This treatment of TTVs inherently neglects time-correlated structure; while this leads to substantial information loss for high-S/N data, low-S/N data can more reasonably be treated in this manner. 

In practice, measurement uncertainties inflate a given planet's observed $V_{i,j}$, which is not considered in Eqn.~\ref{equ:var_no_noise}. To account for this effect, we assume the measured TTV for the $n$th transit of the $i$th planet $\widehat{\text{TTV}}_{i,j}(n)$ is well described by \footnote{In this context, ``$\sim$" denotes ``is distributed as."}
\begin{align}
    \widehat{\text{TTV}}_{i,j}(n) &\sim \mathcal{N} \left(\text{TTV}_{i,j}(n), \sigma_{\text{mid},i,j} \right), 
\end{align}
where  $\sigma_{\text{mid},i,j}$\footnote{We reserve $\sigma$ to denote uncertainties in measuring $\text{TTV}_{i,j}$ and use $V$ to denote variance in the $\text{TTV}_{i,j}$ time-series.} is the uncertainty in the mid-transit time measurement. $\sigma_{{\rm mid},i,j}$ is influenced by the transit S/N, the uncertainty in detrending long-term stellar activity from Kepler light-curves, and the finite integration time of Kepler \citep{Price2014,Holczer2016}. For planet $i$ around star $j$, $\sigma_{{\rm mid},i,j}$ is set equal to the mean of the mid-transit time measurement uncertainties from \cite{Holczer2016}.

Following the above approximations, the population variance for a time-series of TTV observations is $V_{i,j} + \sigma_{\text{mid},i,j}^2$. The sample variance from $N_{i,j}$ transits is\footnote{For $N$ independent draws from the random variable $X$ (indexed by $i=1,\dots,N$), the sample variance is defined as $S^2 = \frac{1}{N-1} \sum_{i=1}^{N} (X_i-\Bar{X})^2$, where $\Bar{X}$ is the mean of the $\{ X_i \}$ samples. }
\begin{align}
    \label{equ:sample_var}
    S^2_{i,j} = S^2_{ \text{planets}, i,j} + S^2_{ \star, i,j} + S^2_{ \text{mid}, i,j}.
\end{align}
The $k$th sample variance component $S^2_{k,i,j}$ is related to the $k$th population variance component via
\begin{align}
    \label{equ:S2_X}
    \frac{(N_{i,j}-1) S^2_{ k, i,j} }{ V_{k,i,j} } &\sim \chi^2_{N_{i,j}-1},
\end{align}
where $k\in\left\{\text{planets},\star, \text{mid} \right\}$. $\chi^2_{N_{i,j}-1}$ is the chi-squared distribution with $N_{i,j}-1$ degrees of freedom. By defining $\beta_{k,i,j} \equiv (N_{i,j}-1) S^2_{k,i,j} /  V_{k,i,j}$ and conducting a change of variables, the $k$th sample variance component is distributed as
\begin{align}
    \label{equ:p(S2_k)}
    p(S^2_{k,i,j} \mid V_{k,i,j}, N_{i,j} ) &= p_{\chi^2_{N_{i,j}-1}}(\beta_{k,i,j}) \Bigg| \frac{d \beta_{k,i,j}}{d S^2_{k,i,j} } \Bigg|, \nonumber \\
     &= p_{\chi^2_{N_{i,j}-1}}(\beta_{k,i,j})  \frac{N_{i,j}-1}{ V_{k,i,j} },
\end{align}
where $p_{\chi^2_{N_{i,j}-1}}(\beta_{k,i,j})$ denotes the $\chi^2_{N_{i,j}-1}$ distribution. The joint probability for the three sample variance components is then
\begin{align}
    \label{equ:p(S2|Sk)}
    p(\{ S_{k,i,j}^2 \} \mid \{ V_{k,i,j} \}, N_{i,j} ) &= \prod_{k} p(S^2_{k,i,j} \mid V_{k,i,j}, N_{i,j} ).
\end{align}
Marginalizing over all $\{ S_{k,i,j}^2 \}$ such that $S^2_{i,j} = \sum_{k} S^2_{k,i,j}$ yields
\begin{align}
    \label{equ:p(S2)}
    p(S^2_{i,j} \mid \{ V_{k,i,j} \}, N_{i,j} ) = & \int_0^{S^2_{i,j}} dX  \int_0^{X} dX^{\prime}  \nonumber \\
    &\times p(S^2_{i,j} - X \mid V_{ \text{planets} ,i,j}, N_{i,j} ) \nonumber \\
    & \times p(X -X^{\prime} \mid V_{\star,i,j}, N_{i,j} ) \nonumber \\
    & \times p(X^{\prime} \mid \sigma^2_{ \text{mid},i,j}, N_{i,j} ),
\end{align}
where the dummy integration variables $X^{\prime}$ and $X$ represent $S^2_{ \text{mid}, i,j}$ and $S^2_{ \text{mid}, i,j}+S^2_{ \star, i,j}$, respectively.

In Eqn.~\ref{equ:p(S2)}, the sample variance $S^2_{i,j}$ of a given planet's TTV measurements is related to that planet's population variance from planet-planet perturbations, stellar noise, and measurement uncertainty. By aggregating a given planet's TTV measurements into a summary statistic, an additional dimension for studying the planet is made available. In Section~\ref{sec:methods_mass}, we use this decomposition of the summary statistic $S^2_{i,j}$ to place constraints on planet mass.

\begin{figure}[t]
\gridline{\fig{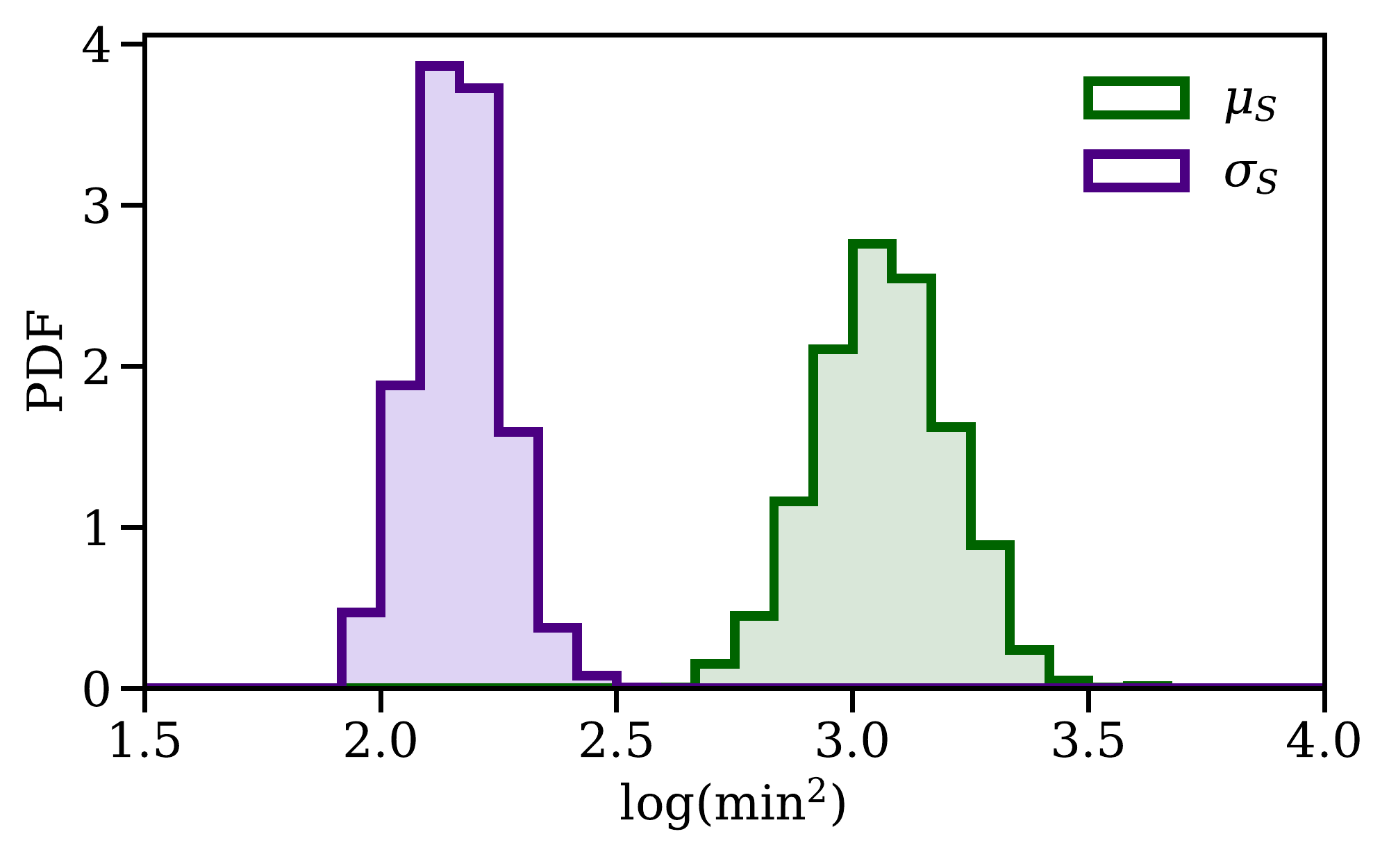}{\columnwidth}{}}
\caption{ Via hierarchical Bayesian inference, the distribution of $V_{\star,i,j}$ for the sample of Kepler systems with only one confirmed planet is well described by a log-normal distribution with $\mu_{\star}=3.08\pm0.14$~log(min$^2$) and $\sigma_{\star}=2.15\pm0.09$~log(min$^2$). In green we present the posterior distribution of $\mu_{\star}$ from MCMC sampling of Eqn.~\ref{equ:p(theta)}, and in purple we present the posterior for $\sigma_{\star}$. }
\label{fig:theta_post}
\end{figure}

\subsection{Underlying distribution of TTVs from stellar noise }
\label{sec:vstar}

Using the above decomposition of $S^2_{i,j}$, planet mass can be constrained by isolating the contribution from $V_{ \text{planets} ,i,j}$. This relies on first inferring the distribution of $V_{\star,i,j}$ for the Kepler sample. To infer the $V_{\star,i,j}$ distribution, we consider the subpopulation of single planet systems. In principle, systems with only one detected planet may harbour several nontransiting planets. However, we assume the subpopulation of single planet systems truly hosts only a single planet (i.e., $V_{ \text{planets},i,j}$ vanishes); see Section~\ref{sec:undetected} for discussion. Under this assumption, a planet's TTV population variance simplifies to a sum of two components, $V_{\star,i,j}+\sigma_{\text{mid},i,j}^2$, allowing the underlying distribution of $V_{\star,i,j}$ to be constrained via hierarchical Bayesian inference. 

Adapting Eqn.~\ref{equ:p(S2)} to the case where $V_{ \text{planets},i,j}$ vanishes, the likelihood of the $j$th single planet system's observed TTV sample variance reduces to
\begin{align}
     p \big(S^2_{1,j} \mid V_{\star,1,j}, & \sigma_{\text{mid},1,j}^2, N_{1,j} \big) = \int_0^{S^2_{1,j}} dS^2_{\text{mid}, 1 ,j} \nonumber \\
     &\times p(S^2_{1,j} - S^2_{ \text{mid},1,j} \mid V_{ \star,1,j}, N_{1,j} ) \nonumber \\
    & \times p(S^2_{ \text{mid},1,j} \mid  \sigma^2_{ \text{mid},1,j}, N_{1,j} ),
\end{align}
where $S^2_{1,j}$ is the sample TTV variance for the one planet in the $j$th system, and $N_{1,j}$ is the number of transits measured. 

For the population distribution of host star induced TTV variance, we consider a log-normal,
\begin{equation}
    p\big(V_{\star,i,j}  \mid \Theta \big) = \frac{1}{ V_{\star,i,j} \sigma_{\star} \sqrt{2\pi}} e^{-(\log V_{\star,i,j}-\mu_{\star})^2/2\sigma_{\star}^2},
    \label{eq:Vstar}
\end{equation}
with the hyperparameters $\Theta=\{ \mu_{\star}, \sigma_{\star} \}$; $\mu_{\star}$ is the mean of the log of the distribution with units of $\log (\rm min^2)$ and $\sigma_{\star}$ is the standard deviation of the log of the distribution with units of $\log (\rm min^2)$. ``log" is reserved to denote the natural logarithm.

We apply a hierarchical modeling approach, using  Eqn.~\ref{eq:Vstar} to set the prior on each individual star's TTV contribution. We denote the set of all the single planets' observed TTV sample variances by $\{ S^2_{1,j} \}$. The likelihood of the $\{ S^2_{1,j} \}$ data set, given the hyperparameters $\Theta$, is then
\begin{align}
    \label{equ:p(theta)}
    p\big(\{ S^2_{1,j}  \}  \mid \Theta, \{ & \sigma_{\text{mid},1,j}^2 \}, \{ N_{1,j} \} \big) = \prod_{j=1}^J \int d V_{\star,1,j} \nonumber \\
    &    \times p \big( V_{\star,1,j} \mid \Theta \big) \nonumber \\
    & \times  p \big(S^2_{1,j} \mid V_{\star,1,j}, \sigma_{\text{mid},1,j}^2, N_{1,j} \big), 
\end{align}
where $j$ indexes over the $J$ single planet systems in the data set.

Via Markov Chain Monte Carlo (MCMC) sampling of Eqn.~\ref{equ:p(theta)},  we generate posteriors for $\mu_{\star}$ and $\sigma_{\star}$ (see Figure~\ref{fig:theta_post}). Sampling is conducted via \texttt{emcee} with 30 walkers for 10,000 iterations, where the first 20\% are rejected as burn-in \citep{emcee2013}; we adopt uniform priors of $\mathcal{U}(0,5)$ and $\mathcal{U}(1,6)$ for $\mu_{\star}$ and  $\sigma_{\star}$, respectively. The posteriors for $\Theta$ are well converged and symmetric, with median values and standard deviations of $\mu_{\star}=3.08\pm0.14$~log(min$^2$) and $\sigma_{\star}=2.15\pm0.09$~log(min$^2$). The posteriors of $\mu_{\star}$ and $\sigma_{\star}$ are uncorrelated, with a Pearson correlation coefficient of $R<0.01$.

The median $\sigma_{ \text{mid},i,j }^2$ for the sample of systems with multiple transiting planets is 202.4~min$^2$, while the median $V_{ \star,i,j }$ from the inferred stellar variance distribution is 22.0~min$^2$. Relative to both the typical $\sigma_{ \text{mid},i,j }$ and the typical transit duration ($>1$~hr), the inferred $V_{ \star,i,j }$ distribution represents a moderate---but nonnegligible---contribution to the measurement of TTVs.

The above procedure is motivated by the expectation that stellar noise sources induce scatter into observed TTV time-series. In practice, the hierarchical model constrains the distribution of excess scatter relative to the reported measurement uncertainties but is agnostic to the origin of the excess scatter. While the distribution of $V_{ \star,i,j}$ inferred above for systems with a single transiting planet likely includes significant contribution from stellar sources, it may also incorporate additional scatter from unaccounted for systematics or misestimation of the measurement uncertainties. The $V_{ \star,i,j}$ distribution is therefore analogous to the stellar jitter white-noise term often invoked in RV modeling \citep[e.g.,][]{Wright2005}.

To identify the primary contributors to the $V_{\star, i,j}$ distribution, several tests could be conducted: (1) inject spot distributions into synthetic transit light-curves and calculate the resulting TTVs to determine what spot parameters (if any) are consistent with the inferred $V_{\star, i,j}$ distribution; (2) isolate potential contributions from stellar noise sources (relative to photometric uncertainties), by investigating the excess TTV scatter as a function of transit S/N and host star properties. These investigations fall beyond the scope of this work but are valuable avenues for future study.

Under the assumption that the $V_{\star,i,j}$ distribution for stars hosting a single planet is consistent with the distribution for stars hosting multiple planets, the inference of $\mu_{\star}$ and $\sigma_{\star}$ can be leveraged to constrain planet mass in multiplanet systems.

\begin{figure*}[t]
\gridline{\fig{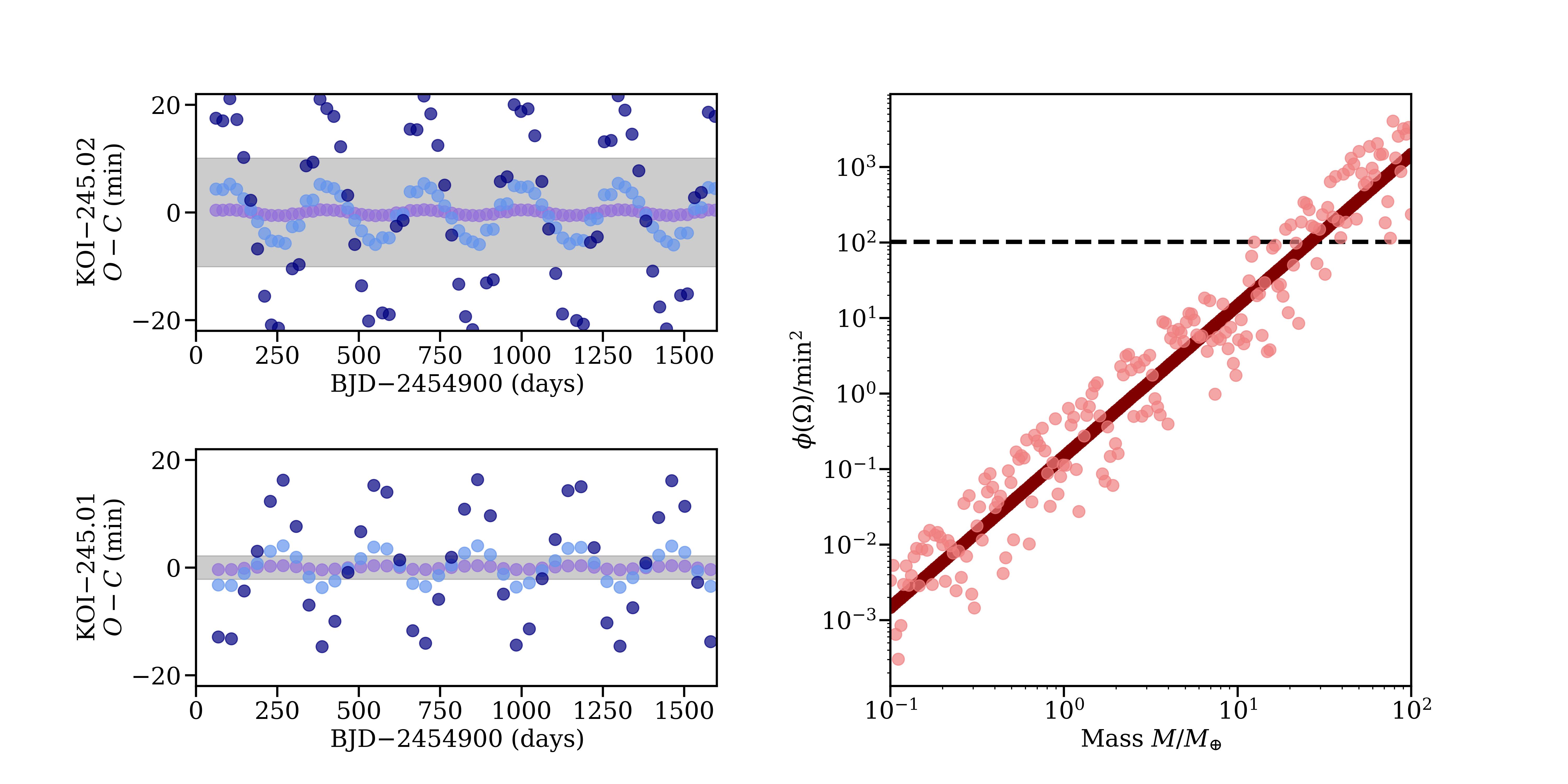}{\textwidth}{}}
\caption{ Here, we diagram our method of deriving mass constraints from low-S/N TTV data. For this illustration, we adopt the orbital periods, initial transit times, and stellar mass of KOI-245 (Kepler-37). In the left column, we present the outer two planets' simulated TTV time-series under three different equal-mass configurations; both planet masses are fixed to either $1~M_{\oplus}$ (purple), $10~M_{\oplus}$ (light blue), or $40~M_{\oplus}$ (dark blue). The TTVs are calculated via \texttt{TTVfaster}. The shaded gray regions demarcate the standard deviation of the observed TTVs (see Figure~\ref{fig:K00245} for the observed TTV time-series). On the right panel, we present $\phi_i(\vec{\Omega}_j)$ for KOI-245.02 (the middle planet) as a function of mass. $\phi_i(\vec{\Omega}_j)$ is the TTV variance for the $i$th planet given a system with dynamical properties $\vec{\Omega}_j$. TTVs are again calculated via \texttt{TTVfaster}. Planet eccentricities are fixed to either zero (dark red) or 0.05 (light red), and the planet masses are again set equal; phases are drawn uniformly between $0$ and $2\pi$. The observed TTV variance for KOI-245.02 is shown as a dashed horizontal line.
}
\label{fig:ttv_diagram}
\end{figure*}

\subsection{Constraining planet mass}
\label{sec:methods_mass}

To place mass constraints on a given planet using low-S/N TTV data, the treatment of $S^2_{i,j}$ from Section~\ref{sec:decomp} must be generalized to multiplanet systems. For a system with $N_{ {\rm p},j}$ planets, the joint probability of observing the sample variances $\{ S_{1,j}^2,\dots,S_{N_{ {\rm p},j},j}^2 \}$ is
\begin{align}
    p\big(\{ & S_{i,j}^2 \} \mid \{ V_{\text{planets},i,j} \},\{ V_{\star,i,j} \},\{ \sigma_{\text{mid},i,j}^2 \}, \{ N_{i,j} \} \big) = \nonumber \\
    & \prod_{i=1}^{N_{ {\rm p},j}} p\left(S^2_{i,j} \mid V_{\text{planets},i,j}, V_{\star,i,j}, \sigma_{\text{mid},i,j}^2, N_{i,j} \right),
\end{align}
under the simplifying approximation that each $S_{i,j}^2$ is independent (i.e., the measurements of the transit times rely on independent segments of the light-curve). 

Marginalizing over the planet-planet perturbation and the stellar noise contributions, and assuming $V_{\star,i,j}$ is uncorrelated between planets, we obtain the likelihood function:
\begin{align}
    \label{equ:long_joint_prob}
    p\big( \{ S_{i,j}^2 & \}  \mid \{ \sigma_{\text{mid},i,j}^2 \},\{ N_{i,j} \}, \Theta, \vec{\Omega}_j \big) = \nonumber \\
    & \prod_{i=1}^{N_{ {\rm p},j}} \int dV_{\text{planets},i,j} \int d V_{\star,i,j} \nonumber \\
    & \times p\big(V_{\text{planets},i,j} \mid \vec{\Omega}_j \big) \times p\left(V_{\star,i,j} \mid \Theta \right)  \nonumber \\
    & \times p\left(S^2_{i,j} \mid V_{\text{planets},i,j}, V_{\star,i,j}, \sigma_{\text{mid},i,j}^2, N_{i,j} \right),
\end{align}
where $\vec{\Omega}_j$ describes a given system's dynamical properties,
\begin{align*}
    \vec{\Omega}_j = \{ &M_1,\dots,M_{N_{ {\rm p},j}}, P_1,\dots,P_{N_{ {\rm p},j}}, \\
    &\varpi_1,\dots,\varpi_{N_{ {\rm p},j}}, e_1,\dots,e_{N_{ {\rm p},j}}, \\
    & T_{1},\dots,T_{N_{ {\rm p},j}} \},
\end{align*}
where $T_{1},\dots,T_{N_{ {\rm p},j}}$ are the initial transit times of the planets according to the mean ephemeris. Following \cite{Agol2016}, mutual inclinations are set to zero. $\vec{\Omega}_j$ and $V_{\text{planets},i,j}$ are related by
\begin{equation}
    p\left(V_{\text{planets},i,j} \mid \vec{\Omega}_j \right) = \delta\left(V_{\text{planets},i,j} - \phi_i(\vec{\Omega}_j) \right).
\end{equation}
$ \delta$ is the Dirac delta function, and $\phi_i(\vec{\Omega}_j)$ is the TTV variance for the $i$th planet calculated from forward modeling of the system for a sufficiently long baseline. To evaluate $\phi_i(\vec{\Omega}_j)$,  we integrate systems for 10,000~days using \texttt{TTVfaster} \citep{Agol2016}. Compared to other methods of TTV calculation, \texttt{TTVfaster} is computationally efficient, is publicly accessible, and scales to higher multiplicities. 

\begin{figure*}[t]
\gridline{\fig{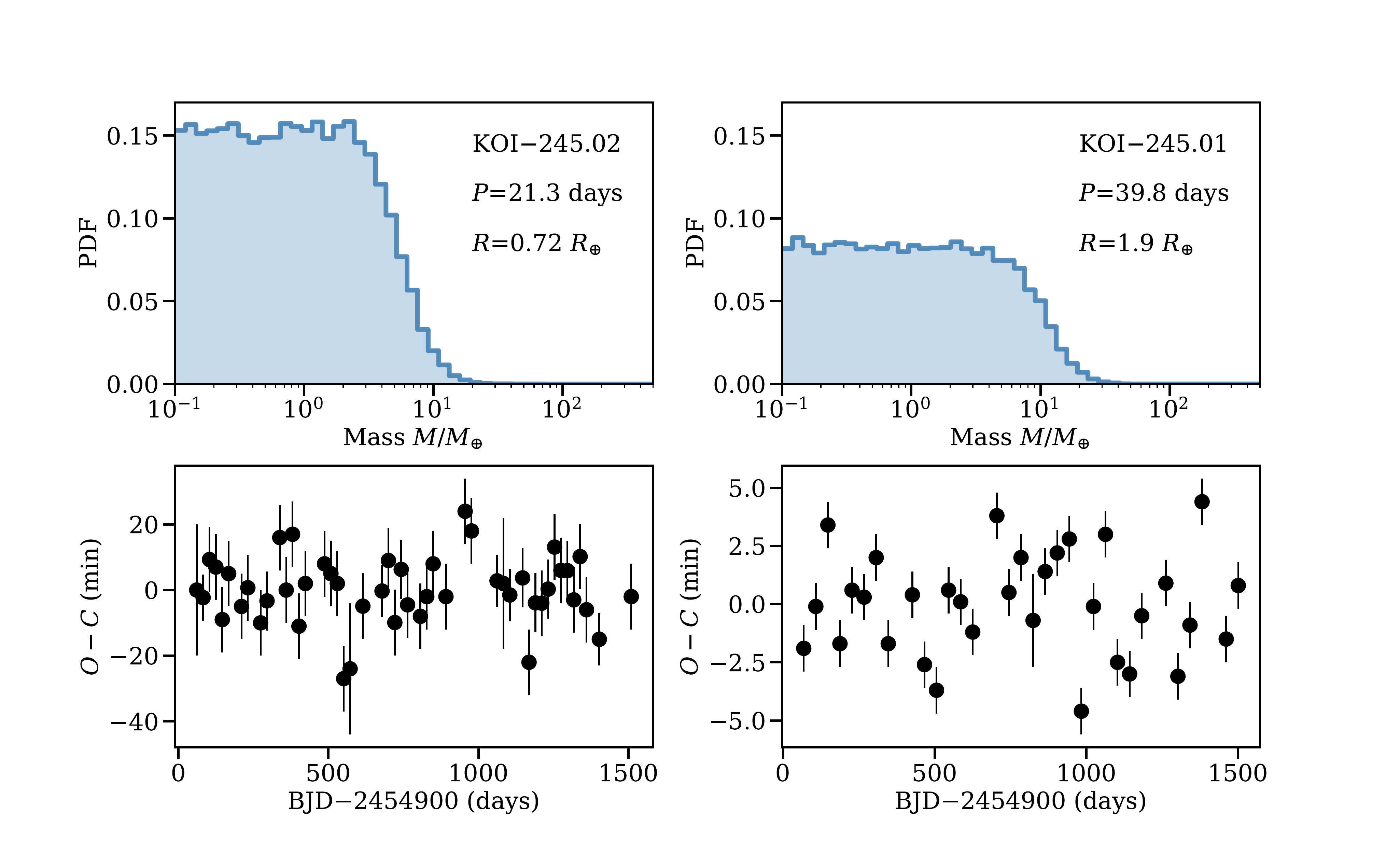}{\textwidth}{}}
\caption{ Using low-S/N TTV data, our method successfully constrains the masses of KOI-245.01 and KOI-245.02 and agrees well with RV-derived mass measurements. On the top row, we present the posteriors on planet mass for KOI-245.02 (left) and KOI-245.01 (right). For reference, we also include the orbital period and radius. On the bottom row, we present the $O-C$ time-series for each KOI. }
\label{fig:K00245}
\end{figure*}

With these approximations, Eqn.~\ref{equ:long_joint_prob} simplifies to
\begin{align}
    \label{equ:short_joint_prob}
    p\big(\{ S_{i,j}^2 & \} \mid  \{ \sigma_{\text{mid},i,j}^2 \},\{ N_{i,j} \}, \Theta, \vec{\Omega}_j \big) = \nonumber \\
    & \prod_{i=1}^{N_{ {\rm p},j}}  \int d V_{\star,i,j} \times   p\left(V_{\star,i,j} \mid \Theta \right) \nonumber \\
    & \times p\left(S^2_{i,j} \mid \phi_i(\vec{\Omega}_j), V_{\star,i,j}, \sigma_{\text{mid},i,j}^2, N_{i,j} \right),
\end{align}
where we have used $V_{\text{planets},i,j}= \phi_i(\vec{\Omega}_j)$.

Assuming $V_{\star,i,j}$ is uncorrelated between planets is an incomplete treatment of the TTV signal. For instance, if the spot coverage on a host star increases, TTV measurements for all planets in the system may see increased noise from spot crossings (assuming the planets are nearly coplanar and the obliquity between the star and planets is near zero, or that the spot coverage on the star is uniform). Increased activity of a host star may also lead to greater residuals in the detrending of long-term stellar activity, resulting in greater scatter in the TTV measurements for all planets in the system.  In Section~\ref{sec:stellar_noise_assumption}, we consider an alternate treatment of stellar noise, in which $V_{\star,i,j}$ is assumed to be the same for planets in a given system.

Posteriors on $\vec{\Omega}_j$ are obtained by weighting Monte Carlo samples of $\vec{\Omega}_j$ with respect to Eqn.~\ref{equ:short_joint_prob}. For a given system, planet properties are independently drawn from a Rayleigh prior distribution on eccentricity \citep[scale factor of 0.02 unless otherwise stated;][]{He2019}, a uniform prior distribution on $\varpi$ from $0$ to $2\pi$, and a uniform prior distribution on log$_{10}$-mass from $0.1$~$M_{\oplus}$ to $1000$~$M_{\oplus}$. Adopting a uniform prior on eccentricity from either $0-0.025$ or $0-0.05$, instead of the Rayleigh prior, has minimal effect on the posteriors. Mutual inclinations are fixed to zero. The Keplerian orbital periods and the initial transit times are fixed to the observed values. Two-hundred thousand samples are drawn for each system.

To diagram these methods, we consider the three-planet system KOI-245 (Kepler-37). Kepler-37 hosts three small planets ($R<2~R_{\oplus}$) with orbital periods of 13, 21, and 40 days \citep{Barclay2013}. Given its small size ($R=0.3~R_{\oplus}$), the innermost planet is omitted from the \cite{Holczer2016} TTV catalog. This omission has a minimal effect on the mass constraints, because the small, innermost planet likely perturbs the outer planets very weakly. Unlike systems typically characterized via TTVs, the TTVs for KOI-245.01 and KOI-245.02 show minimal time-correlated  structure and have excess scatter ratios well below those typical of systems with TTV-based mass constraints. KOI-245.01 and KOI-245.02 have $S^2_{i,j}$ of 5.0 and 104.5 min$^2$ and $\sigma_{ \text{mid}, i,j}^2$ of 1.1 and 105.5 min$^2$, respectively.

For three different mass configurations, we present the TTV time-series calculated by \texttt{TTVfaster} for the outer two planets (Figure~\ref{fig:ttv_diagram}). The planets are placed on flat, circular orbits with equal-masses, fixed to either $1,10,$ or $ 40~M_{\oplus}$. Qualitatively, planet masses $\gtrsim10~M_{\oplus}$ are disfavored relative to lower masses, because the expected TTV signal for these higher masses greatly surpasses the observed TTV scatter. The relation between TTV scatter and planet mass (for an equal-mass configuration) is presented in the right panel of Figure~\ref{fig:ttv_diagram}.

\begin{figure}[t]
\gridline{\fig{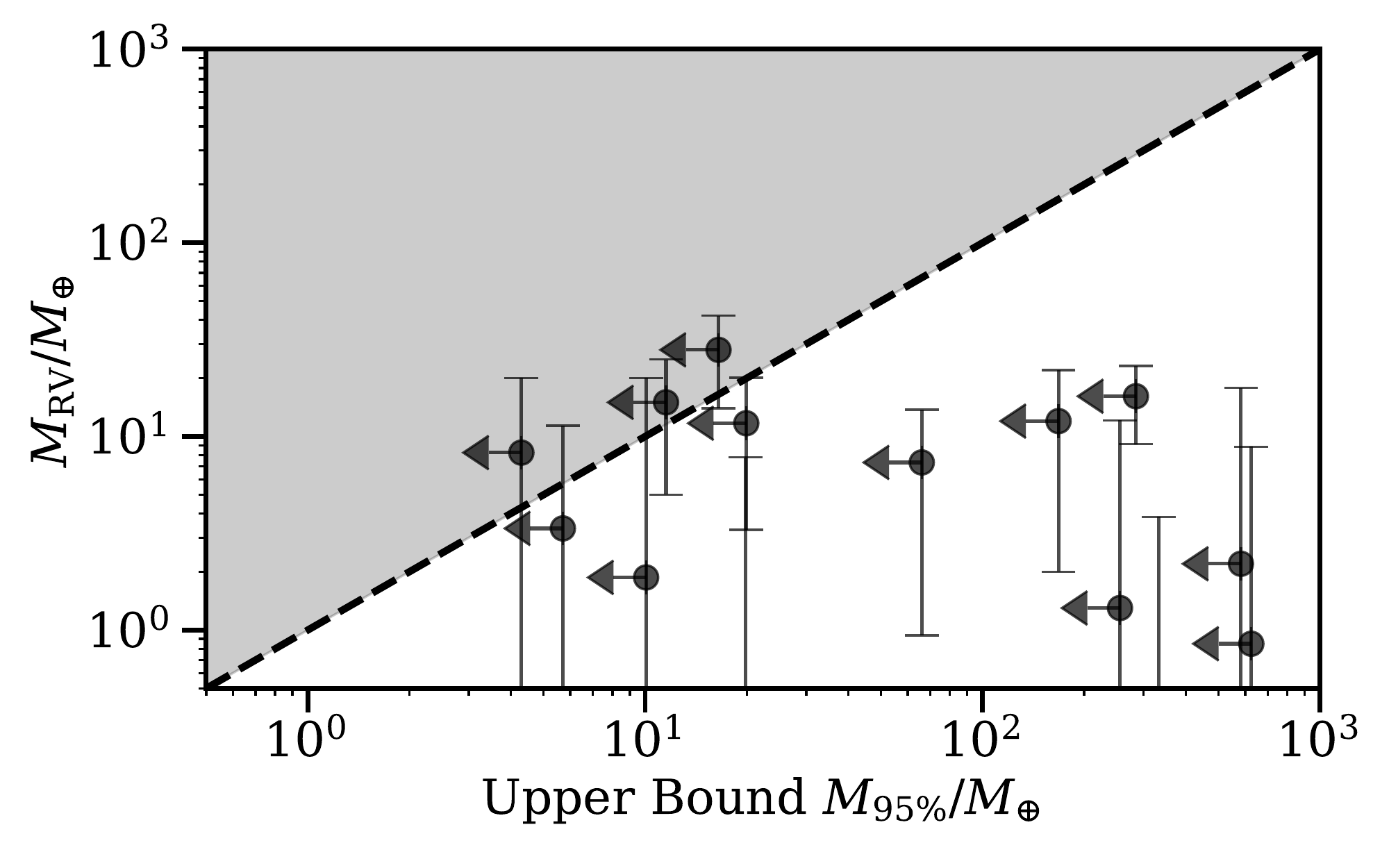}{\columnwidth}{}}
\caption{ Low-S/N TTV-derived upper bounds on planet mass are consistent with RV-derived mass measurements. Here, we present the $95$th percentile on planet mass from our methods against RV-derived masses for the 14 planets in our sample with RV masses. The vertical error bars show the 2$\sigma$ uncertainties on the RV measurements. }
\label{fig:mrv_m95}
\end{figure}

Following the procedure outlined above, multiplanet systems can be characterized using existing low-S/N TTV data. In Section~\ref{sec:results}, we apply this method to 175~planets in 79~systems (see Section~\ref{sec:sample} for discussion of the sample) and discuss the resulting mass constraints.

\section{Results}
\label{sec:results}

Following the methods outlined in Section~\ref{sec:methods}, we calculate posteriors on planet mass for a sample of Kepler planets using low-S/N TTVs. The sample consists of 175~planets (in 79~systems) included in both the \cite{Holczer2016} TTV catalog and the CKS sample; in principle, these methods can be applied to any multiplanet system with TTV measurements. 

As a diagnostic example, we present the mass constraints and TTV time-series for KOI-245 (Kepler-37, described in Section~\ref{sec:methods_mass}) in Figure~\ref{fig:K00245}. Despite the low-S/N of the TTVs, our method places meaningful constraints on the system and strongly disfavors planet masses $ \gtrsim 10~M_{\oplus}$ for KOI-245.01 and KOI-245.02; using our standard $V_{\star,i,j}$ treatment, low-S/N TTVs set upper bounds ($95$th percentile of the posteriors) of $5.7$ and $10.0~M_{\oplus}$ for KOI-245.02 and KOI-245.01, respectively.

Independent mass estimates for KOI-245 are limited. RV measurements yielded mass upper bounds of $12.0$ and $12.2~M_{\oplus}$ for KOI-245.02 and KOI-245.01, respectively \citep{PSCompPars}. There is no expectation that the upper bounds from low-S/N TTVs lie near the upper bounds from RV measurements---only that the mass posterior from RV measurements does not lie significantly above the low-S/N TTV bound; the similarity between the RV and low-S/N TTV upper bounds for KOI-245 is coincidental. \cite{Hadden2014} reported a mass estimate of $\sim63~M_{\oplus}$ for KOI-245.01, based on the apparent perturbations KOI-245.01 induced on the adjacent planet candidate KOI-245.04; however, KOI-245.04 has since been flagged as a false positive, invalidating this mass constraint for KOI-245.01.

\begin{figure}[t]
\gridline{\fig{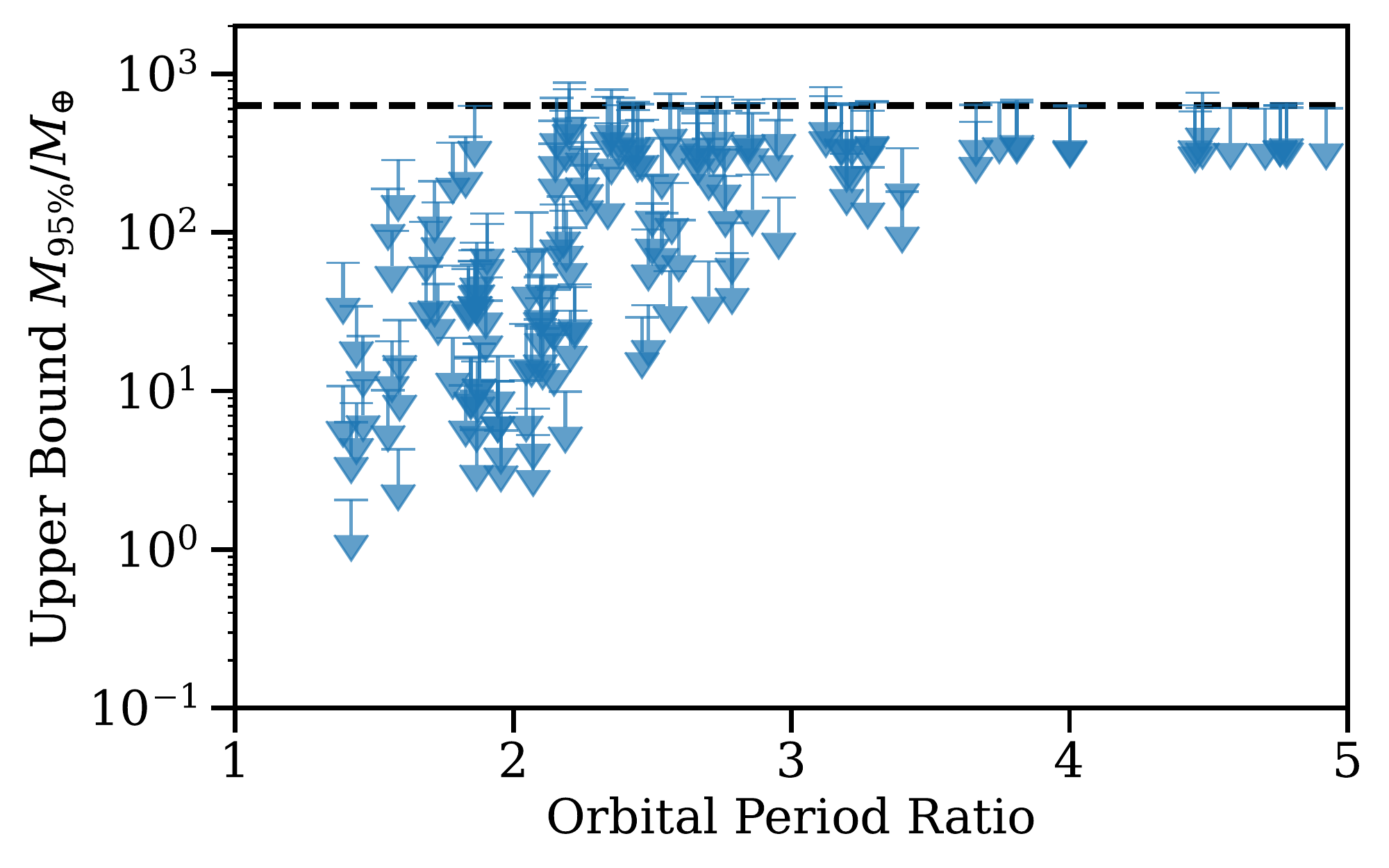}{\columnwidth}{}}
\caption{ The strength of a planet's mass constraint is generally correlated with the planet's minimum adjacent period ratio. Here, we present each planet's TTV-derived upper bound on planet mass (parameterized by the $95$th percentile of the posterior, $M_{95\%}$) alongside the planet's minimum adjacent period ratio. The $95$th percentile of our planet mass prior is shown as a dashed horizontal line. }
\label{fig:period_ratio}
\end{figure}

\begin{figure}[t]
\gridline{\fig{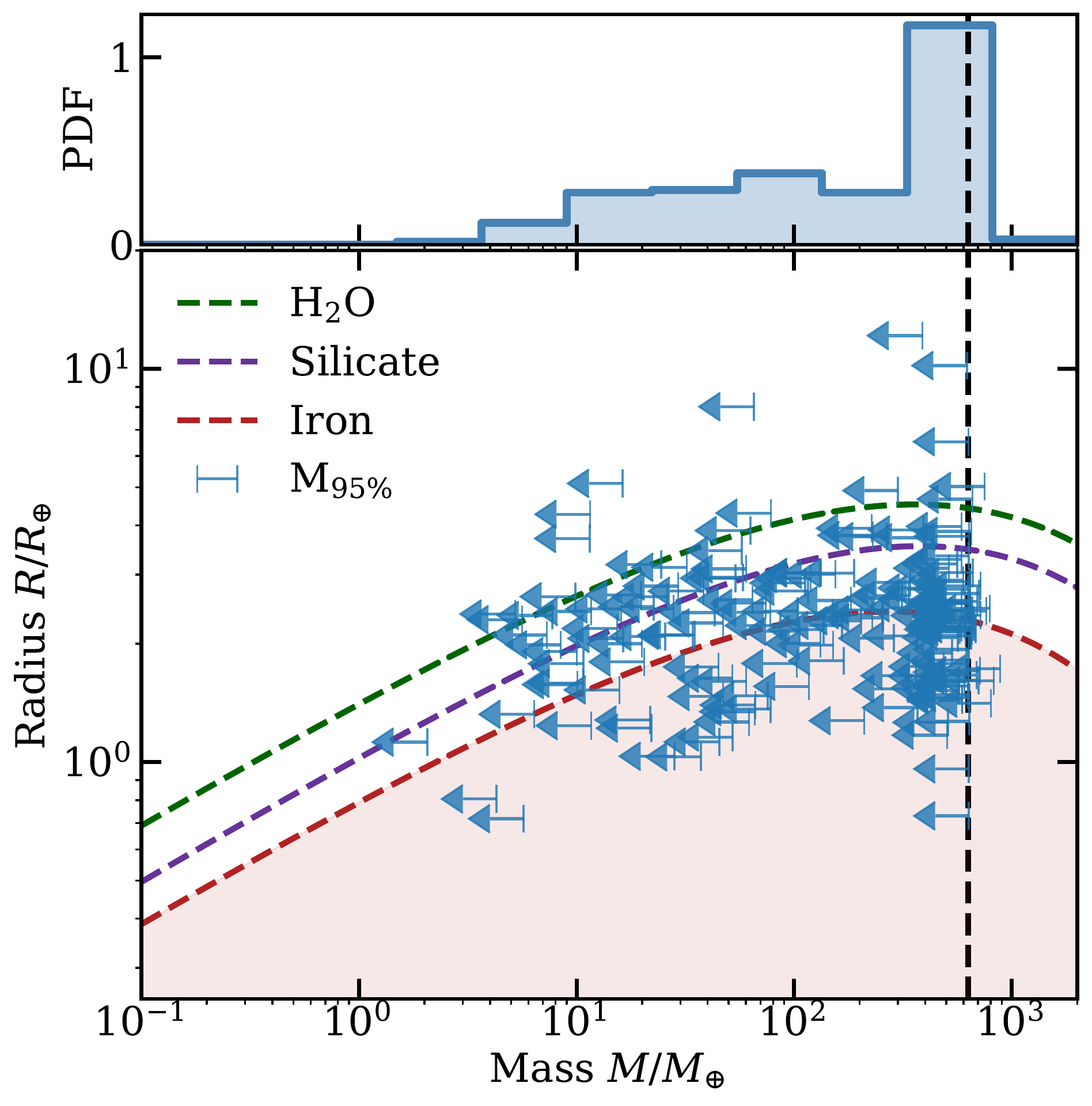}{\columnwidth}{}}
\caption{ For $>50$ planets, our methods place informative constraints on planet mass (i.e., the $95$th percentile on planet mass  $M_{95\%}$ significantly deviates from the prior and yields a bulk density less than a pure iron composition). We present $M_{95\%}$ for each planet's posterior against its radius (blue), alongside the mass-radius relation for a pure iron (red), silicate (purple), and water-ice (green) composition \citep{Seager2007}. The $95$th percentile of our mass prior is shown as a dashed vertical black line. }
\label{fig:m95}
\end{figure}

For a more robust diagnostic of the newly derived mass constraints, we consider the subsample of planets with RV-derived masses. Including KOI-245.01 and KOI-245.02, there are 14 planets in 6 systems within our sample that have RV-derived mass measurements or mass upper limits. In Figure~\ref{fig:mrv_m95}, we present each planet's RV-derived mass alongside the $95$th percentile on mass from the low-S/N TTV posteriors, denoted $M_{95\%}$. As expected, $M_{95\%}$ is greater than the RV-derived mass (within uncertainties) for each KOI.

For three planets---KOI-137.01, KOI-137.02, and KOI-283.02---the central mass estimate from independent measurements falls above the low-S/N TTV upper bound. While the RV mass estimate for KOI-283.02 has considerable uncertainty (consistent with zero at $2\sigma$), the RV mass posteriors for KOI-137.01 and KOI-137.02 are relatively narrow. A joint RV-TTV fit to KOI-137 (Kepler-18) yielded mass estimates of $17.3\pm1.9$ and $16.4\pm1.4~M_{\oplus}$ \citep{Cochran2011}, while low-S/N TTVs yielded upper bounds of $11.5$ and $16.5~M_{\oplus}$ for KOI-137.01 and KOI-137.02, respectively. Since KOI-137.01 and KOI-137.02 both present clear quasi-sinusoidal TTVs, the adoption of a uniform prior of planet phase potentially led to overly stringent low-S/N TTV-derived mass constraints.

Compared to the median or mean, $M_{95\%}$ is a preferred summary statistic of the mass posteriors. For low-S/N TTV data, where $S^2_{i,j} \approx \sigma_{\text{mid},i,j}^2$, the observed sample variance is already near the expected value, and any significant planet-planet perturbations are strongly disfavored. For this reason, the mass posteriors nominally flatten at lower planet masses, because those samples each yield negligible  $V_{ \text{planets},i,j }$, and strongly taper at higher masses. This trend is clearly seen in the posteriors for KOI-245 (Figure~\ref{fig:K00245}). The mass posteriors are therefore more readily interpreted in terms of upper bounds. 

If a given KOI's observed $S^2_{i,j}$ is significantly greater than its $\sigma_{\text{mid},i,j}^2$ and the typical contribution from $V_{\star,i,j}$, a nonnegligible contribution from planet-planet perturbations is necessary. The TTV likelihood function (Eqn.~\ref{equ:short_joint_prob}) may therefore peak at larger planet masses and even favor extreme mass ratios for the system, depending on the $S^2_{i,j}$ and $\sigma_{\text{mid},i,j}^2$ of the other planets. However, for such systems, time-series modeling of the TTVs is likely preferable, to take advantage of the higher TTV S/N. 

Upper bounds on mass ($M_{95\%}$) and composition constraints for all 175 planets in the sample are publicly available; a sample is provided in Table~\ref{tab:upperbounds}.

Given the nature of TTVs, we expect the strength of a planet's mass constraint to scale with the planet's period ratio (i.e., stronger constraints at lower period ratios and closer to resonances). As an additional validation test,  we compare each planet's $M_{95\%}$ with its minimum adjacent period ratio (Figure~\ref{fig:period_ratio}). As expected, $M_{95\%}$ tends to deviate from the prior at lower period ratios, with strong deviations near the 2:1 resonance. At period ratios above three, the mass constraints are generally consistent with the prior. The fraction of planets with significant mass constraints (i.e., deviating from the prior) is therefore closely tied to the period ratio distribution of the selected sample. For the Kepler sample, the period ratio distribution peaks between $1.5$ and $2$ and decays steadily at higher values \citep{Fabrycky2014,Brakensiek2016}. The methods outlined in Section~\ref{sec:methods} are most sensitive to the most common systems in the Kepler sample.   

\begin{deluxetable*}{ccccccccc}[t]
\tablecaption{ Upper Bounds on Planet Mass \label{tab:upperbounds}}
\tablehead{
 \colhead{KOI} & \colhead{Radius} & \colhead{Alt. $M_{95\%}$ } &   \colhead{Std. $M_{95\%}$ } & \colhead{Deviates from Prior} & \colhead{Bulk Comp.} \\
  \colhead{} &  \colhead{($R_{\oplus}$)} & \colhead{($M_{\oplus}$)} & \colhead{($M_{\oplus}$) } & \colhead{}
}
\startdata
41.02 & 1.34 & 30.1 & 65.9 & \checkmark & ---\\
41.01 & 2.34 & 945.1 & 627.1 & --- & ---\\
41.03 & 1.54 & 171.4 & 333.1 & --- & ---\\
85.02 & 1.52 & 578.1 & 563.9 & --- & ---\\
85.01 & 2.58 & 76.0 & 64.2 & \checkmark & $<\rho_\mathrm{iron}$\\
85.03 & 1.78 & 10.8 & 10.8 & \checkmark & $<\rho_\mathrm{iron}$\\
111.01 & 2.38 & 5.1 & 5.3 & \checkmark & $<\rho_\mathrm{water}$\\
111.02 & 2.36 & 9.9 & 7.7 & \checkmark & $<\rho_\mathrm{silicate}$\\
111.03 & 2.64 & 10.0 & 9.9 & \checkmark & $<\rho_\mathrm{silicate}$\\
123.01 & 2.33 & 254.3 & 256.5 & --- & ---\\
123.02 & 2.54 & 713.3 & 586.0 & --- & ---\\
\dots & \dots & \dots & \dots & \dots & \dots
\enddata
\tablecomments{ For each KOI, the $95\%$ percentile on planet mass from the low-S/N TTV-derived posteriors is reported using the standard treatment of stellar noise---uncorrelated between planets in a given system---and using the alternate noise treatment---identical between planets in a given system. Bounds on bulk composition are obtained by comparing each planet's least constraining upper bound on mass, i.e., whichever $M_{95\%}$ is higher between the standard and alternative stellar noise treatments, with the mass-radius relations of \cite{Seager2007}. The notation ``$<\rho_X$'' denotes that $M_{95\%}<M_X(R)$, where $M_X(R)$ is the mass of a planet of radius $R$ composed of $X$; we consider $X \in \{ \text{water-ice, silicate, iron} \}$.  A null composition constraint is reported if $M_{95\%} > M_\mathrm{Fe}(R)$ or the mass upper bound does not significantly deviate from the prior on mass (i.e., $M_{95\%} > 200~M_{\oplus}$). If a planet's radius is greater than the maximum radius of a pure water-ice planet ($\sim 4.5~R_{\oplus}$), the bulk composition is already constrained to be highly gaseous, regardless of the low-S/N TTV mass bound. For planets where $M_{95\%} < 200~M_{\oplus}$ and $R > 4.5~R_{\oplus}$, the bulk composition is marked with an asterisk. Planet radii are adopted from \cite{Fulton2018}. This table is published in its entirety in machine-readable format. A portion is shown here for guidance regarding its form and content. }

\end{deluxetable*}

For a significant fraction of planets ($>30\%$ in our sample), low-S/N TTVs place informative constraints on planetary composition. In Figure~\ref{fig:m95}, we present $M_{95\%}$ for each planet alongside the planets' radii. For a planet of radius $R$, the low-S/N TTV-derived mass constraint is considered informative if $M_{95\%}<M_\mathrm{Fe}(R)$, where $M_\mathrm{Fe}(R)$ is the mass of a pure iron planet with radius $R$, and $M_{95\%}$ diverges from the prior on mass (i.e., $M_{95\%} < 200~M_{\oplus}$). Within our sample, 53 planets satisfy this criteria. 

Of the planets with informative mass constraints, low-S/N TTVs favor volatile-rich compositions for $>20$ planets. Similar to the selection of informative mass constraints, planets are considered likely volatile-rich if $M_{95\%}<M_\mathrm{silicate}(R)$, where $M_\mathrm{silicate}(R)$ is the mass of a pure silicate planet with radius $R$, and $M_{95\%}$ diverges from the prior on mass; 25 planets satisfy this criteria. The mass constraints for eight planets are even strong enough to satisfy $M_{95\%}<M_\mathrm{water}(R)$. The mass-radius relations for pure iron, silicate, and water-ice compositions are adopted from \cite{Seager2007}. For a significant fraction of the sample, low-S/N TTVs successfully constrain the planets' bulk compositions.

KOI-2194.01 and KOI-2194.02---two small planets of radius $2.3$ and $2.1~R_{\oplus}$, respectively---are of particular interest. Upon inspection, KOI-2194.02 presents quasi-sinusoidal TTV variation with an amplitude of $\sim50$~minutes; KOI-2194.02 was not included in the list of KOIs with significant long-term TTVs by \cite{Holczer2016}. Low-S/N TTVs constrain the maximum bulk density of these planets to approximately a pure water-ice composition. Further study of the TTVs in this system, along with greater investigation of the apparent low-bulk densities of the two planets, is warranted.

\begin{figure}[t!]
\gridline{\fig{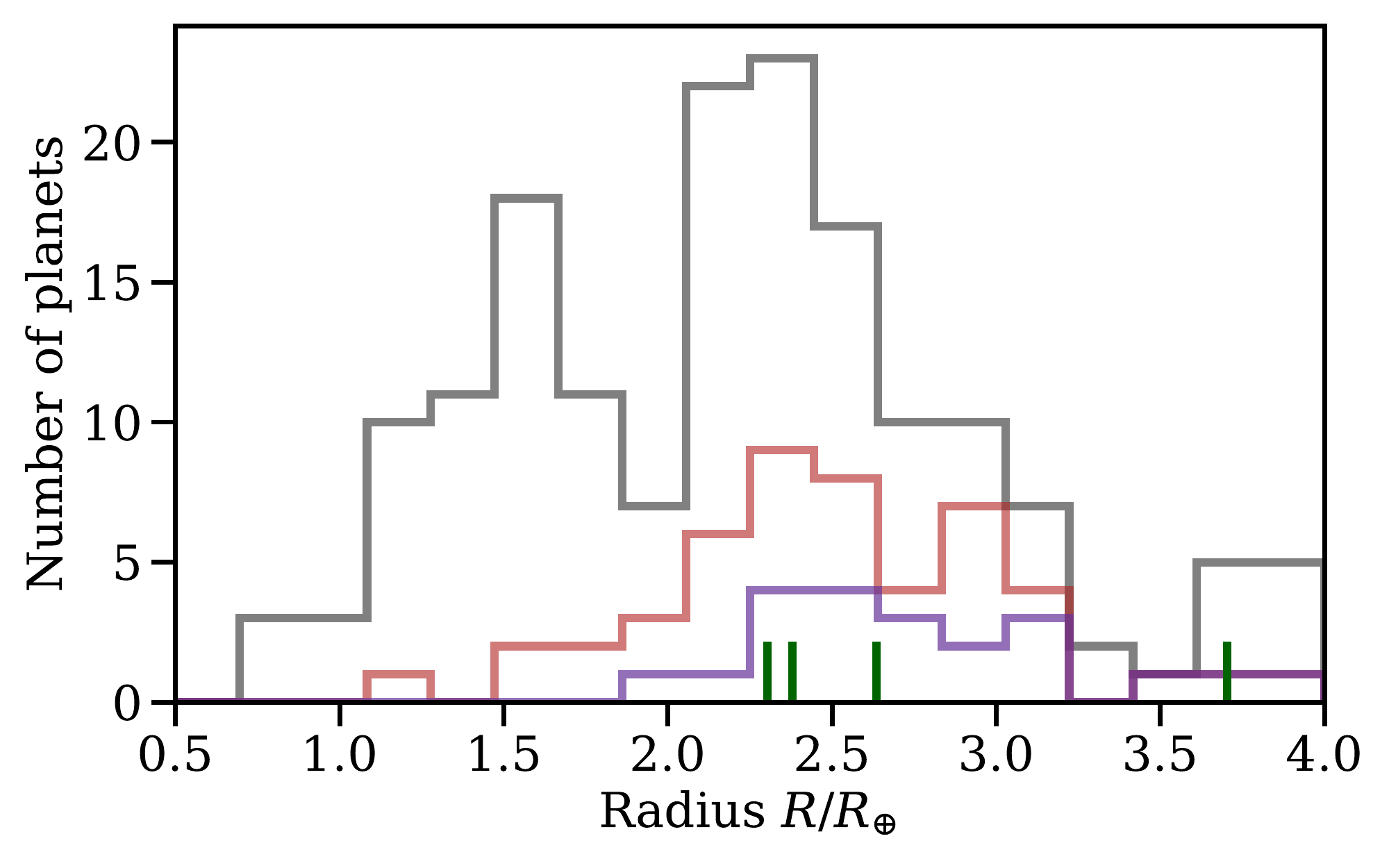}{\columnwidth}{}}
\caption{Planets with low-S/N TTV-derived composition constraints are predominately larger than $2~R_{\oplus}$. Here, we present the planet radius distribution for our entire sample of planets (black), alongside the distributions for planets with informative mass constraints (red) and planets with informative mass constraints that are less dense than a pure silicate composition (purple). Planets with informative mass constraints that are less dense than a pure water-ice composition are presented as vertical green lines. For clarity, the radius distributions are limited to $R < 4~R_{\oplus}$. }
\label{fig:composition_hist}
\end{figure}

The planets where low-S/N TTVs most strongly constrain planetary composition are predominately larger than $2~R_{\oplus}$. This trend reflects a bias of the low-S/N TTV constraints. Since the mass posteriors are most appropriately leveraged as upper bounds, and the mass-radius relations for pure iron, silicate, and water-ice compositions are positively sloped for $R\lesssim100~R_{\oplus}$, ruling out a rocky composition for a smaller planet requires a more stringent mass constraint than for a larger planet. This trend is evident in Figure~\ref{fig:composition_hist}, where we present the planet radius distribution for our entire sample of planets alongside three subpopulations: (i) all planets with informative low-S/N TTV mass constraints, (ii) planets with informative mass constraints that are less dense than a pure silicate composition, and (iii) planets with informative mass constraints that are less dense than a pure water-ice composition.

Using low-S/N TTVs, 25 planets have mass upper bounds that demand volatile-rich compositions. These limits represent a substantial increase in the population of planets with composition constraints. These planets are therefore of great interest for future study. In total, low-S/N TTV mass constraints yield informative upper bounds on planet mass for 53 planets ($>30\%$ of the sample). 

\section{Discussion}
\label{sec:disc}

Here, we first investigate the sensitivity of our methods to the approximations made in Section~\ref{sec:methods}, including the effect of undetected planets (Section~\ref{sec:undetected}) and the treatment of stellar noise (Section~\ref{sec:stellar_noise_assumption}).  In Section~\ref{sec:correlations}, we then briefly consider planet-planet correlations in the posteriors, and in Section~\ref{sec:demographics}, we discuss the application of low-S/N TTV-based mass constraints to future studies of planetary composition and demographics.

\subsection{Effect of undetected planets}
\label{sec:undetected}

The methods outlined in Section~\ref{sec:methods} assume nondetected planets contribute negligibly---compared to noise from the host star---to the observed variance in a given planet's TTV time-series. In particular, in Section~\ref{sec:vstar} we employ hierarchical Bayesian modeling to characterize the excess noise in the TTV time-series of single planet systems. Assuming the contribution from nondetected planets is negligible, we associate this excess signal with stellar noise. However, the occurrence rate and architectures of nondetected planets remains a point of ongoing study \citep[e.g.,][]{Lissauer2011,Hansen2013,Ballard2016,Millholland2021}. We therefore adopt a population-synthesis approach to assess the validity of our assumption.

\begin{figure}[t!]
\gridline{\fig{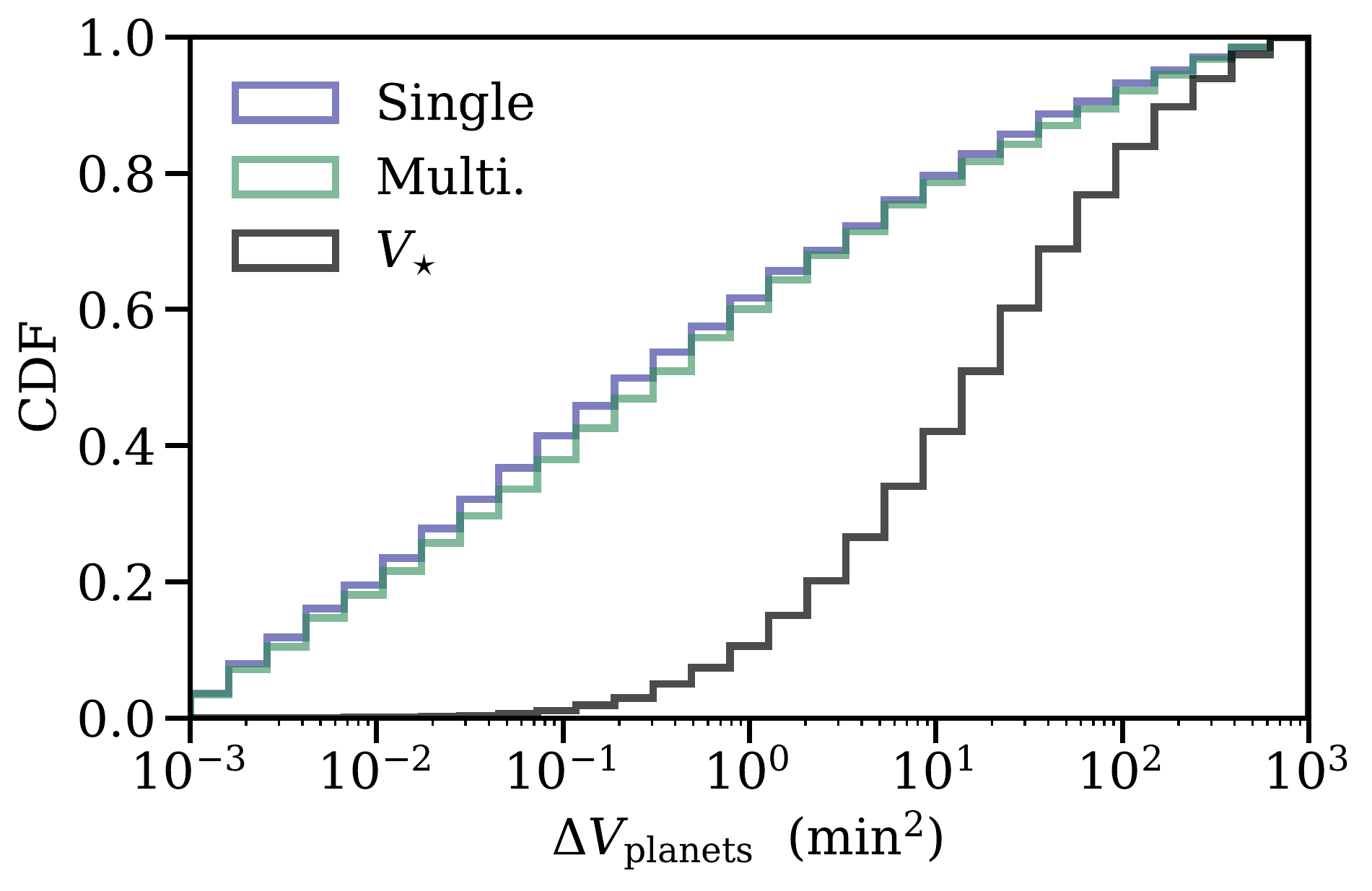}{\columnwidth}{}}
\caption{ For the vast majority of synthetic systems from \cite{He2020}, the TTV variance induced by nondetected planets is significantly lower than the excess noise inferred for the Kepler singles. Here, we present the difference in TTV variance for a synthetic Kepler sample between modeling the physical and observed systems from  \cite{He2020}; systems where only one mock planet is observed are shown in blue. For comparison, we present the excess noise component---characterized via hierarchical Bayesian modeling in Section~\ref{sec:vstar}---for the Kepler singles in black.  }
\label{fig:missing_planets}
\end{figure}

Population-synthesis models \citep[e.g., ][]{Hsu2018,Mulders2018,He2019,He2020} generate a synthetic population of stars and planetary systems and then convolve these systems with detection effects to yield a simulated sample of observed planets. In addition to constraining the underlying distributions of planet size and architectures, forward models enable the study of \textit{physical} planetary systems, as opposed to \textit{observed} planetary systems. We leverage population-synthesis to investigate the TTV time-series of a synthetic planet when either all planets in the system are modeled or only the detected planets are modeled. By comparing the two cases for a synthetic Kepler sample, the impact of nondetected planets on TTVs can be constrained. 

For this analysis, we consider the synthetic Kepler catalogs of the Exoplanet System Simulator \citep[SysSim;][]{He2020}. In this model, planets' orbital periods and radii are drawn from a series of underlying distributions, while the planets' eccentricities and mutual inclinations are assigned such that the system has an angular momentum deficit (AMD) at the critical value for stability. AMD is defined as the difference between the total angular momentum of the system and the angular momentum of the system if all orbits are circular and flat \citep{Laskar1997}. Via approximate Bayesian computing optimization of the hyperparameters, the model successfully replicates key observed features of the Kepler sample, including the Kepler dichotomy and peas-in-a-pod trends \citep{He2020}. By assigning mutual inclinations via the maximum AMD method, the systems of \cite{He2020} resolve the Kepler dichotomy with a single population of systems---as opposed to having a population of intrinsic singles and a population of multiplanet systems---suggesting most systems with only one detected planet in fact harbour several additional nondetected planets. The maximum AMD SysSim catalogs therefore offer a conservative assessment of the impact of nondetected planets in systems with a single detected transiting planet. 

Using 250,000 systems from the maximum AMD SysSim catalogs, we model both the \textit{physical} and corresponding \textit{observed} systems using \texttt{TTVfaster}. In Figure~\ref{fig:missing_planets}, we present the distribution of $\Delta V_\mathrm{planets}$, the difference in TTV variance between modeling the physical and observed systems, against the distribution of excess noise inferred for the host stars of Kepler singles (derived in Section~\ref{sec:vstar}). To parallel the filter on excess scatter ratio applied in Section~\ref{sec:sample}, we only consider synthetic planets if the TTV variance (for both the physical system and the observed system) divided by the mid-transit time measurement uncertainty is less than 5. Analogous to \cite{Steffen2016}, each planet is assigned a characteristic mid-transit time measurement uncertainty by taking the average measurement uncertainty of the 10 planets in the \cite{Holczer2016} catalog closest in radius to the synthetic planet. 

Even using the maximum AMD systems from \cite{He2020}---for which many observed single planet systems host additional nondetected planets---the inferred contribution from stellar noise among Kepler singles is considerably higher than the TTV variance induced by nondetected planets. The $75$th percentile of the inferred $V_{\star, i,j}$ distribution is over 7 times the $75$th percentile of the $\Delta V_\mathrm{planets}$ distribution for systems with only one detected planet; this falls to a factor of 1.5 at the $95$th percentile. 

For the vast majority of single planet systems, population-synthesis suggests the excess signal in the TTV time-series does not originate from nondetected planets; however, for a small fraction of systems ($\sim 5\%$), nondetected planets may produce a nonnegligible signal. The assumption that the TTV signal induced by nondetected planets is negligible relative to stellar noise is therefore reasonable for our analysis.

\subsection{Alternate treatment of stellar noise}
\label{sec:stellar_noise_assumption}

\begin{figure}[t!]
\gridline{\fig{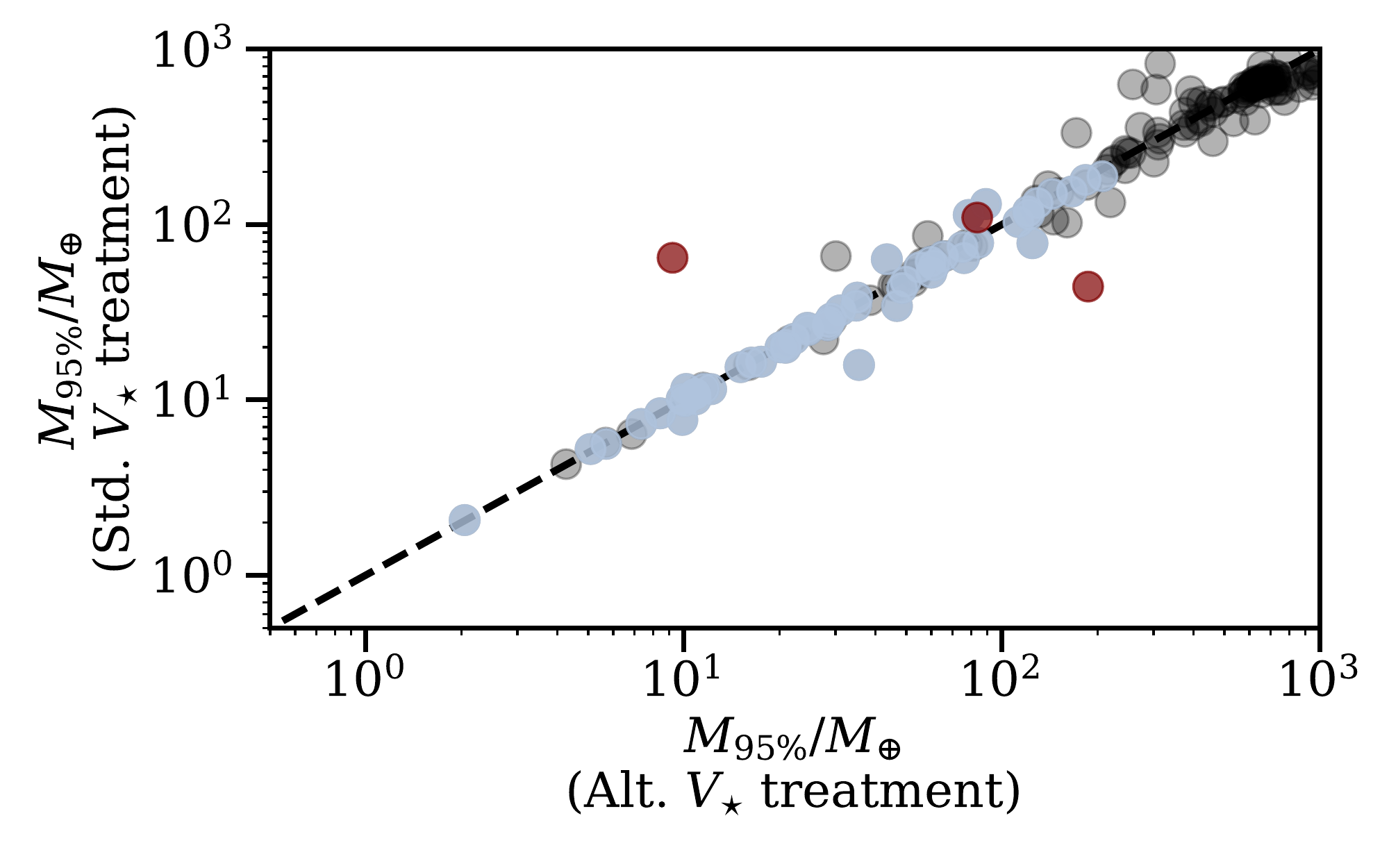}{\columnwidth}{}}
\caption{ Low-S/N TTV-derived mass constraints are generally insensitive to the treatment of stellar noise $V_{\star,i,j}$. Here, we present the upper bounds on planet mass $M_{95\%}$ using our standard $V_{\star,i,j}$ treatment (i.e., assuming $V_{\star,i,j}$ is uncorrelated between planets) against the upper bounds derived using our alternative $V_{\star,i,j}$ treatment (i.e., assuming $V_{\star,i,j}$ is identical for planets in a given system). Planets where our standard assumption yields informative mass constraints are highlighted in blue. The KOI-658 system, where the two treatments differ by more than a factor of 4 for the outer two planets, is highlighted in red.}
\label{fig:vstar_comp}
\end{figure}

In Section~\ref{sec:methods}, the TTV variance of a given planet was decomposed into three sources: noise from measurement uncertainties ($\sigma_{ \text{mid} ,i,j}$), stellar noise ($V_{\star,i,j}$), and planet-planet perturbations ($V_{ \text{planets} ,i,j}$). The distribution of $V_{\star,i,j}$ was then inferred via hierarchical modeling of the sample of systems with only one detected transiting planet. In practice, the hierarchical model inferred the distribution of excess scatter relative to the reported measurement uncertainties. Therefore, it may also incorporate additional scatter from unaccounted for systematics or misestimation of the measurement uncertainties, in addition to stellar noise sources.

In deriving the TTV likelihood function (Eqn.~\ref{equ:short_joint_prob}), we adopted the simplifying assumption that $V_{\star,i,j}$ is uncorrelated between planets in a given system. However, this treatment is incomplete. Stellar noise sources (e.g., spot crossings, photometric variability from granulation, residuals in long-term light-curve detrending, etc.) are inherently linked to the activity level and properties of the host star. Within a given system, each planet's $V_{\star,i,j}$ may be correlated with the other planets.  While the details of intrasystem $V_{\star,i,j}$ correlations are beyond the scope of this work, here we consider an alternate treatment of $V_{\star,i,j}$ to explore the impact of this assumption on our results. 

\begin{figure*}[t!]
\gridline{\fig{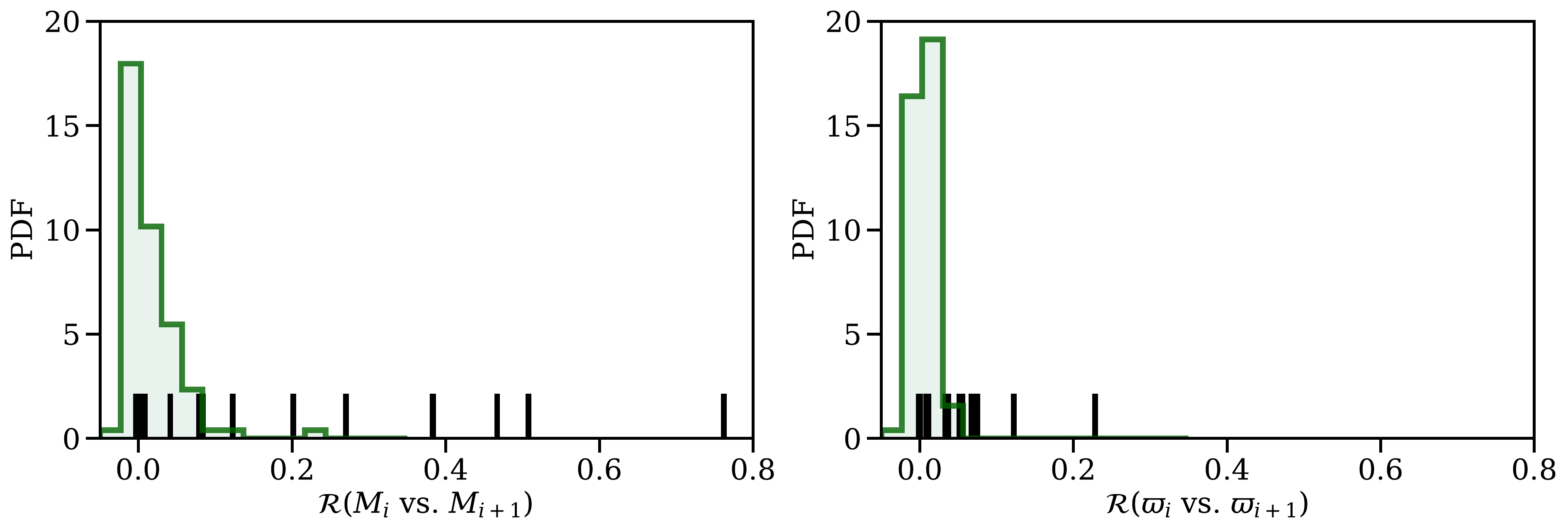}{\textwidth}{}}
\caption{ Every planet in our sample shows minimal planet-planet correlations for both the mass and longitude of periastron posteriors. We present the distribution of the Pearson correlation coefficient $\mathcal{R}$ between adjacent planets' mass posteriors (left panel) and the correlation coefficient between adjacent planets' longitudes of periastron (right panel). The planet-planet correlations for the near resonant systems removed from our sample are shown as black vertical lines. }
\label{fig:correlations}
\end{figure*}

Rather than assume $V_{\star,i,j}$ is uncorrelated between planets in a given system, we may make the converse assumption: $V_{\star,i,j}$ is identical for all planets in a system (i.e., $V_{\star,j} \equiv V_{\star,i,j}$). Under this alternate assumption, the TTV likelihood function is
\begin{align}
    \label{equ:short_joint_prob_alt}
    p\big(\{ S_{i,j}^2 & \} \mid  \{ \sigma_{\text{mid},i,j}^2 \},\{ N_{i,j} \}, \Theta, \vec{\Omega}_j \big) = \nonumber \\
    & \int d V_{\star,j} \times p\left(V_{\star,j} \mid \Theta \right)   \nonumber \\
    & \times  \prod_{i=1}^{N_{ {\rm p},j}} p\left(S^2_{i,j} \mid \phi_i(\vec{\Omega}_j), V_{\star,j}, \sigma_{\text{mid},i,j}^2, N_{i,j} \right).
\end{align}
Eqn.~\ref{equ:short_joint_prob_alt} is analogous to Eqn.~\ref{equ:short_joint_prob}, where $V_{\star,i,j}$ is assumed to be uncorrelated between planets in a given system. As in Eqn.~\ref{equ:short_joint_prob}, we substitute $V_{\text{planets},i,j}= \phi_i(\vec{\Omega}_j)$.

Following the Monte Carlo procedure outlined in Section~\ref{sec:methods_mass}, we generate posteriors for each multiplanet system in our sample using Eqn.~\ref{equ:short_joint_prob_alt} as the likelihood function. In Figure~\ref{fig:vstar_comp}, we present the upper bounds on planet mass $M_{95\%}$ derived using our standard treatment of $V_{\star,i,j}$ against the upper bounds derived using our alternate treatment. With the exception of the KOI-658 system (Kepler-203), the low-S/N TTV-derived mass constraints show minimal dependence on the adopted stellar noise treatment; excluding Kepler-203, the upper bounds change by less than a factor of 2.7. 

Kepler-203 hosts three $2$--$3~R_{\oplus}$ sized planets with orbital periods of 3.2, 5.4, and 11.3 days. The TTVs of the inner two planets have excess scatter ratios near 1, and the outer planet has an excess scatter ratio near 2. While the mass constraint for the inner planet is largely insensitive to the adopted stellar noise treatment, the upper bounds on mass for the outer two planets change by more than a factor of 4 between the two likelihood functions. If the stellar noise component is assumed to be identical among planets in a system, high masses are invoked for the middle planet to induce large TTVs for the outer planet. However, when $V_{\star,i,j}$ is assumed to be uncorrelated between planets, high masses are disfavored for the middle planet; the outer planet can be assigned a greater stellar noise component than the inner two planets and no significant planet-planet perturbation signal is necessary. The intrasystem diversity of excess scatter ratio in the Kepler-203 system therefore exacerbates the differences between the two $V_{\star,i,j}$ treatments.

For the vast majority of systems, the low-S/N TTV-derived mass constraints are insensitive to the adopted stellar noise treatment.

\begin{figure*}[t!]
\gridline{\fig{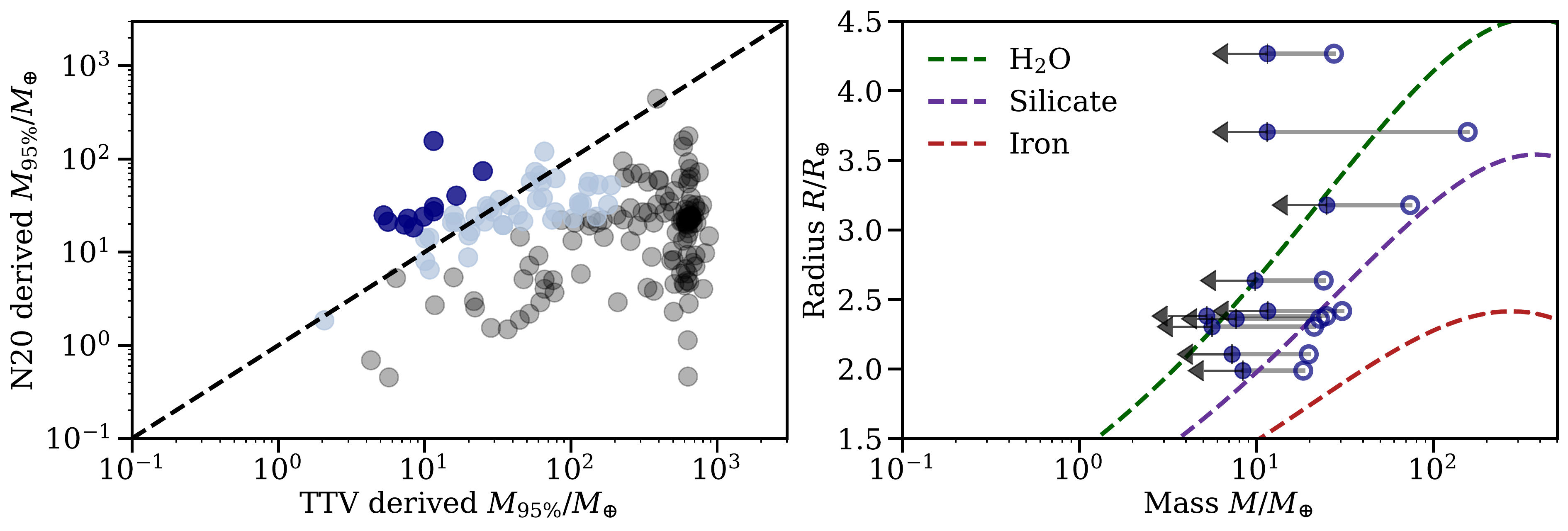}{\textwidth}{}}
\caption{ Masses derived from low-S/N TTVs offer valuable constraints on the demographics of Kepler planets. On the left, we present the $95$th percentile on planet mass from low S/N TTVs against the $95$th percentile on planet mass from the joint mass-radius-period distribution  of \cite{Neil2020}; see Section~\ref{sec:demographics} for discussion. Planets where the TTV-derived mass constraint is informative (i.e.,  $M_{95\%}<M_\mathrm{Fe}(R)$, and  $M_{95\%}$ deviates from the prior on mass) are shaded light-blue; planets with informative TTV mass constraints that are a factor of 2 lower than the $95$th percentile constraint from \cite{Neil2020} are shaded dark-blue. On the right, we present the difference in $M_{95\%}$ between our TTV-based method (solid blue points) and \citep[][hollow blue points]{Neil2020}, for the planets where this change is greater than a factor of two. To contextualize these changes, we present the mass-radius relations for pure iron (red), silicate (purple), and water-ice (green) compositions \citep{Seager2007}.  }
\label{fig:demographics}
\end{figure*}
\subsection{Planet-planet correlations}
\label{sec:correlations}

The likelihood function derived in Section~\ref{sec:methods_mass} for low-S/N TTVs (Eqn.~\ref{equ:short_joint_prob}) considers the TTVs of every planet in a given system, as opposed to treating planets individually. The posteriors derived for each system could therefore exhibit strong planet-planet correlations. 

For high-S/N TTVs, model posteriors often show strong planet-planet correlations, particularly between planet phases \citep[e.g.,][]{JontofHutter2016}. However, since our method only considers TTV amplitude, correlations between adjacent planets' longitudes of periastron likely indicate the system is near resonant, and that the system goes unstable for certain longitude configurations. As discussed in Section~\ref{sec:sample}, near resonant systems are removed from our sample, because they often require fine-tuned orbital elements, such as the need for convergent migration to achieve long-term stability in the TRAPPIST-1 system \citep{Tamayo2017}. We therefore expect our sample of planets to show minimal planet-planet correlations between longitudes of periastron.

For both planet mass and longitude of periastron, adjacent planets in our sample show minimal planet-planet correlations. In Figure~\ref{fig:correlations}, we present the distribution of the Pearson correlation coefficient $\mathcal{R}$ between adjacent planets' mass posteriors, as well as the distribution for the phase posteriors. As expected, the 11 systems removed from our sample for being near resonant show greater planet-planet correlations than our nominal sample. 

Every planet in our sample shows minimal planet-planet correlations between their posteriors. The sample filtering process (Section~\ref{sec:sample}) therefore successfully flagged near resonant systems that are incompatible with the methods of Section~\ref{sec:methods}. 

\subsection{Demographics studies}
\label{sec:demographics}

Constraints on planet mass are critical to studies of planetary formation and evolution. In particular, such studies rely on mass constraints precise enough to distinguish between various planetary compositions. As discussed in Section~\ref{sec:results}, for 53 planets ($>30\%$ of our sample) low-S/N TTV-based mass constraints yield informative upper bounds on planet mass, i.e., $M_{95\%}<M_\mathrm{Fe}(R)$, where $M_\mathrm{Fe}(R)$ is the mass of a planet of radius $R$ composed entirely of iron, and $M_{95\%}$ deviates from the prior on mass ($M_{95\%} < 200~M_{\oplus}$). Through comparison to single-composition equations-of-state from \cite{Seager2007}, TTV-based mass constraints can even rule out rocky or icy compositions for several planets (see Figure~\ref{fig:m95}). 

To further establish the diagnostic power of low-S/N TTV-based mass constraints and to motivate future study, we next consider the application of our methods to demographics studies of the Kepler sample. 

Using available mass constraints for the Kepler sample, prior studies have attempted to constrain the radius-mass plane for small planets \citep[e.g., ][]{Weiss2013,Wu2013}. However, investigations into the radius gap have highlighted the importance of constraining the joint mass-radius-period distribution of planets \citep[e.g., ][]{Fulton2017,Fulton2018,vanEylen2018,Gupta2019}. Recently, \cite{Neil2020} investigated the joint mass-radius-period distribution of Kepler planets via hierarchical Bayesian mixture models, with XUV-driven hydrodynamic mass loss; the model was trained on the CKS sample of planets, using Gaia cross-matched radii and masses derived from RV measurements where available (68/1130 planets). A mixture model with gaseous envelope, evaporated core, and intrinsically rocky planet populations was preferred by the data. Given the hierarchical nature of the model, \cite{Neil2020} inferred the mass and radius of each planet in the CKS sample, in addition to the hyperparameters on the underlying  planet property distributions. By comparing the predicted planet masses from \cite{Neil2020} with our TTV-derived mass constraints, we can highlight how strongly our methods will inform demographics studies; in particular, significant disagreement with the predicted masses suggests TTV-derived mass constraints offer previously unavailable diagnostic power.

For each planet in our sample, we compare the $95$th percentile on planet mass from our TTV methods with the $95$th percentile on mass from \cite{Neil2020}; see the left panel of Figure~\ref{fig:demographics}. Of the planets where low S/N TTV-based mass constraints are informative, the upper bounds on planet mass are consistent with \cite{Neil2020} for the majority of planets, i.e., $M_{95\%}$ from \cite{Neil2020} is less than our TTV-derived upper bound. For planets with informative low S/N TTV-constraints, our upper bounds are also positively correlated with the upper bounds of \cite{Neil2020},  as expected. However, for eleven systems, $M_{95\%}$ from \cite{Neil2020} is more than a factor of two greater than the TTV-derived upper bound. For several of these planets, including low S/N TTV constraints moves the upper bound on mass from a rocky or icy composition to a gaseous composition; see the right panel of Figure~\ref{fig:demographics}.

Mass constraints derived from low-S/N TTVs have great potential to advance demographics studies of the Kepler sample; such constraints not only are accessible using publicly available data and are informative for a substantial fraction of planets ($>30\%$ of our sample) but are also capable of distinguishing between different planetary compositions. 

\section{Conclusions}
\label{sec:conc}

Prior studies inferred planet mass from TTVs \citep[e.g.,][]{Xie2013,Hadden2014,Xie2014,JontofHutter2016,Hadden2017} by modeling TTVs as a function of time using either numerical integration \citep{Deck2014}, first-order in eccentricity and mass analytic solutions \citep{Agol2016}, or a linear combination of basis functions \citep{Hadden2016,Linial2018,Hadden2019}. However, such methods are restricted to high-S/N TTVs. 

Here, we developed and implemented a method of constraining planet mass in multiplanet systems using low-S/N TTVs from Kepler. We aggregated a given TTV time-series into a summary statistic---variance---which was then decomposed into three independent components: noise from measurement uncertainties, noise from the host star, and planet-planet perturbations. Through a series of simplifying approximations, we derived a likelihood function for the observed TTV variances of the planets in a system given a set of system properties, e.g., masses, eccentricities, phases, and orbital periods. By constraining the distribution of stellar noise in the Kepler sample via hierarchical Bayesian analysis and leveraging the likelihood function derived in Section~\ref{sec:methods_mass} (Eqn.~\ref{equ:short_joint_prob}), upper bounds on planet mass were inferred for each planet in the sample. 

Low-S/N TTV-derived upper bounds on planet mass are consistent with RV mass measurements (for the subset of planets where such measurements are available). The mass posteriors show minimal dependence on the assumption that nondetected planets induce negligible TTV signal compared to noise from host stars and show minimal dependence on the adopted treatment of stellar noise.

Of our sample of 175 planets in 79 multiplanet systems, low-S/N TTVs yield informative upper bounds on planet mass for 53 planets ($>30\%$ of the sample); the mass constraint for a planet of radius $R$ is considered informative if the $95$th percentile on planet mass is less than the mass of a planet of radius $R$ composed entirely of iron, and the upper bound on mass strongly deviates from the prior. For 25 planets, low-S/N TTVs favor volatile-rich compositions. These results represent a significant increase in the sample of planets with composition constraints.

Upper bounds on planet mass and corresponding composition constraints are made publicly available for every planet in the sample (see Table~\ref{tab:upperbounds}).

We anticipate low-S/N TTVs will provide a powerful path toward expanding the sample of Kepler planets with mass constraints. Since TTV time-series are available for each Kepler planet, these methods are scalable to every multiplanet Kepler system with low-S/N TTVs; using our limited sample, low-S/N TTVs have already yielded meaningful mass constraints for $>50$~planets. Low-S/N TTV-derived mass constraints have also proven capable of distinguishing between various planetary compositions. Such mass constraints should therefore offer significant diagnostic power to future demographics studies.  

\ \\
\acknowledgments
We thank Jacob Bean, Daniel Fabrycky, Sam Hadden, and Samuel Halverson for helpful conversations. This research has made use of the NASA Exoplanet Archive, which is operated by the California Institute of Technology, under contract with the National Aeronautics and Space Administration under the Exoplanet Exploration Program. This work was also completed in part with resources provided by the University of Chicago’s Research Computing Center. L.A.R. gratefully acknowledges support from NASA Habitable Worlds Research Program grant 80NSSC19K0314, from NSF FY2016 AAG Solicitation 12-589 award number 1615089, and the Research Corporation for Science Advancement through a Cottrell Scholar Award.

\bibliography{paper}%

\end{document}